\documentclass[11pt]{article}

\usepackage{amsmath}
\usepackage{amssymb}
\usepackage{mathtools}
\usepackage{mathrsfs}
\usepackage{amsfonts}
\usepackage{bbm}
\usepackage{enumerate}
\usepackage{graphicx}
\usepackage{array}
\usepackage{booktabs}
\usepackage{textcomp}
\usepackage{caption}
\usepackage{cite}
\usepackage{cancel}
\usepackage{units}
\usepackage{multirow}
\usepackage{pst-all}
\usepackage{longtable}
\usepackage{relsize}

\newcommand{\zseven}[0]{\ensuremath{\mathbbm{Z}_{7}}}
\newcommand{\ee}[0]{\ensuremath{E_8}}
\newcommand{\eexee}[0]{\ensuremath{E_8 \times E_8}}
\newcommand{\SU}[1]{\ensuremath{SU(#1)}}
\newcommand{\SO}[1]{\ensuremath{SO(#1)}}
\newcommand{\mmod}[0]{\ensuremath{\text{ mod }}}
\newcommand{\tr}[0]{\ensuremath{\text{tr}}}

\begin{document}

\thispagestyle{empty}
~\vspace{0.5cm}
\begin{center}

{\Large {\bf A perfect match of MSSM--like orbifold and resolution models via anomalies}}\\[0pt]

\bigskip
\bigskip {\large
{\bf Michael Blaszczyk$^{a,}$}\footnote{
E-mail: michael@th.physik.uni-bonn.de},
{\bf Nana Geraldine Cabo Bizet$^{a b,}$}\footnote{
E-mail: nana@th.physik.uni-bonn.de},
{\bf Hans Peter Nilles$^{a,}$}\footnote{
E-mail: nilles@th.physik.uni-bonn.de},
{\bf Fabian~Ruehle$^{a,}$}\footnote{
E-mail: ruehle@th.physik.uni-bonn.de}
\bigskip }\\[0pt]
\vspace{0.23cm}
${}^{a}${\it Bethe Center for Theoretical Physics,\\
~~Physikalisches Institut der Universit\"at Bonn, 
Nussallee 12, 53115 Bonn, Germany\\[12pt]
${}^{b}$ Centro de Aplicaciones Tecnol\'{o}gicas y Desarrollo Nuclear,\\
Calle 30, esq.a 5ta Ave, Miramar, La Habana, Cuba
}
\bigskip
\end{center}

\subsection*{\centering Abstract}

Compactification of the heterotic string on toroidal orbifolds is a promising set--up for the construction of realistic unified models of particle physics. The target space dynamics of such models, however, drives them slightly away from the orbifold point in moduli space. This resolves curvature singularities, but makes the string computations very difficult. On these smooth manifolds  we have to rely on an effective supergravity approximation in the large volume limit. By comparing an orbifold example with its blow--up version, we try to transfer the computational power of the orbifold to the smooth manifold. Using local properties, we establish a perfect map of the the chiral spectra as well as the (local) anomalies of these models. A key element in this discussion is the Green--Schwarz anomaly polynomial. It allows us to identify those redefinitions of chiral fields and localized axions in the blow--up process which are relevant for the interactions (such as Yukawa--couplings) in the model on the smooth space.

\vskip 1cm
\clearpage
\thispagestyle{empty}
\tableofcontents
\clearpage
\setcounter{page}{1}
\setcounter{footnote}{0}

\section{Introduction}
\label{sec:Introduction}

Orbifold compactification \cite{Dixon:1985jw,Dixon:1986jc,Ibanez:1986tp} of the heterotic string \cite{Gross:1984dd,Gross:1985fr} is a promising approach towards the construction of supersymmetric extension of the standard model (MSSM) and grand unified theories \cite{Georgi:1974sy}. It leads to a plethora of discrete symmetries \cite{Kobayashi:2006wq} which turn out to be very useful for phenomenological considerations in the MSSM, such as proton stability \cite{Forste:2010pf,Lee:2010gv}, flavor universality \cite{Ko:2007dz}, suppression of the $\mu$--term \cite{Casas:1992mk,Antoniadis:1994hg,Lebedev:2007hv} and the creation of the hierarchy between the GUT-- and the weak scale \cite{Kappl:2008ie}. Interactions in orbifold compactifications are fully calculable in the framework of conformal field theory \cite{Dixon:1986qv,Hamidi:1986vh} (other approaches are Gepner models \cite{Dijkstra:2004cc,Dijkstra:2004ym,GatoRivera:2009yt} and free fermionic constructions \cite{Faraggi:1989ka,Kiritsis:2008ry}). This gives us full control over stringy effects in these consistent UV--completions of the MSSM. Many of the properties of the models are determined through the localization of the fields in extra dimensions on fixed points and fixed tori \cite{Forste:2004ie,Kobayashi:2004ya,Buchmuller:2004hv,Buchmuller:2005jr}, leading to new insights in model building such as e.g.\ the concept of local grand unification \cite{Nilles:2009yd,Nilles:2008gq,Buchmuller:2005sh}.

Still, orbifold compactification represents a specific point in the moduli space of smooth (Calabi--Yau) compactification and it is not clear how close nature is located to those fixed points and/or fixed tori. In fact, at the orbifold point the models give rise to many exotic states as well as additional $U(1)$ gauge bosons. It is also a point of enhanced discrete symmetries and not all of these are exactly realized in nature. Some of them need to be (slightly) broken creating a hierarchy of scales as e.g. useful for a pattern of quark and lepton masses through a Froggatt--Nielsen machanism \cite{Froggatt:1978nt}.

We thus would not expect to sit exactly at the fixed points, but still should not be too far away. Consistent constructions of potentially realistic models incorporate the seed for such a mechanism in form of an anomalous $U(1)$ that drives the theory away from the fixed point. It induces a non--trivial Fayet--Iliopoulos (FI) term that leads to a breakdown of superfluous $U(1)$s and allows the removal of exotic states \cite{Atick:1987gy,Dine:1987xk,Font:1988mm}. The analysis of this mechanism cannot be done up to now in string theory itself, but has to rely on methods in the low--energy effective supergravity (SUGRA) theories which are believed to be reliably close to the fixed points.

Still there remains the question whether there is a more stringy way to understand this mechanism. Of course, ideally we would like to use a full--fledged Calabi--Yau compactification of the heterotic string \cite{Donagi:1999ez,Donagi:2000zs,Braun:2005ux,Braun:2005nv}, but there we can (due to the complexity of the manifolds) only compute a limited number of properties. Central tools are index theorems valid for generic points in moduli space and they do not, in general, capture the richness of special points in moduli space that might be relevant for realistic model building. We thus need a way to connect orbifolds to smooth manifolds in a (locally) controlled manner, keeping the powerful computational tools from orbifold compactification.

It is here where the so--called blow--up mechanism \cite{Candelas:1985en,Aspinwall:1994ev,Nibbelink:2007rd,Nibbelink:2007pn} could play a crucial role. It opens new geometrical insight, although the computational techniques are less powerful. With a precise map from the orbifold-- to the blow--up model we could, however, still rely on calculations in the orbifold limit. Such a precise map would require an ``exact'' match of the spectrum to make sure that the orbifold blow--up pair is correctly identified.

Such an analysis is the main goal of this paper. In previous attempts to blow up realistic orbifold models \cite{Nibbelink:2009sp,Blaszczyk:2010db} as e.g.\ models of the $\mathbbm{Z}_{6-II}$ Mini--Landscape \cite{Lebedev:2006kn,Lebedev:2007hv,Lebedev:2008un} (another realistic orbifold construction has been considered in \cite{Kim:2007mt}) or more recently in $\mathbbm{Z}_2 \times \mathbbm{Z}_2$ models \cite{Blaszczyk:2009in} there were some obstructions to identify such precise maps. Ambiguities in the spectrum arising from flop transitions \cite{Blaszczyk:2010db} or ``brother models'' from discrete torsion \cite{Vafa:1994rv,Ploger:2007iq} enter the discussion and leave some questions. It is not clear whether this is just due to the complexity of the models or whether there is a general obstruction. We need further inspections to identify the central elements of the precise match. To approach this question we study here a somewhat simplified version based on the \zseven\ orbifold. This model \cite{Casas:1990yt} has the complexity of the Mini--Landscape models (including realistic gauge group) but avoids some subtleties that will have to be clarified later.

One of the main results of this paper is the construction of a precise match for such a  pair of models. Another even more important result is the observation that (local) anomalies play a crucial role in the search for this match. On the orbifold, localized anomalies \cite{Gmeiner:2002es} can be understood via the localization of chiral states at the various fixed points, a picture that becomes obscured in the resolved version. It is here that the Green--Schwarz (GS) anomaly polynomial \cite{Green:1984sg,Schellekens:1986xh} starts to play a central role. On the orbifold the Green--Schwarz anomaly cancellation involves a unique axion field and is thus pretty simple: the information about localized anomalies can be explicitly seen in the spectrum itself. The blow--up version, however, contains many axions and therefore the GS anomaly cancellation becomes more subtle. This is the key point in the transition from the orbifold to the blow--up version: local properties which are obvious from the orbifold point of view become apparent in the blow--up version through the non--trivial structure of the GS anomaly polynomial. Many of the properties in the orbifold models are encoded in the anomaly polynomial and can be used to read off local information in the resolved model that would otherwise not be available in the blown up version.

The outline of the paper is as follows. In section 2 we present the \zseven\ orbifold model and its candidate blow--up partner. Section 3 is devoted to the study of the spectra of the pair of models and the matching of states. This includes a choice of field redefinitions (motivated by the anomaly considerations) and the definition of a local multiplicity operator in the blow--up version which is more powerful than global index theorems on smooth manifolds. Equipped with this tool we can then analyze the matching of states. We identify states that appear massless on the orbifold side and massive in blow--up and vice versa and explain the origin of this mismatch in the naive calculation. A perfect match between the pair of models is therefore achieved.

Section 4 is devoted to the study of the anomaly polynomial and its role in blow--up. We identify the non--universal (localized) axions and their appearance in the anomaly polynomial. This allows us to identify the anomalies in the resolved space and discuss their crucial role in the blow--up process. We give a detailed discussion of the various anomalies and identify the blow--up modes as non--universal axions. Many of the properties of the model are encoded in the ``anomalies'' such as the mixing of blow--up modes (which is important for Yukawa couplings in the blow--up model). This concludes the ``perfect match'' between the pair of models and shows that a detailed study of the match reveals important information on the properties of the model. Section 5 is devoted to some concluding remarks and some indication for future research along this direction.

\section{Orbifold and resolution models} 

One theory we are working with is the compactification of the heterotic string on a six dimensional toroidal orbifold. We briefly review the geometrical properties, in particular the fixed point structure which is crucial for the low energy spectrum of such a setup. The second theory is compactifying ten dimensional heterotic SUGRA on the smooth resolution of the $\mathbbm{Z}_7$ orbifold. Since stringy effects are surpressed by powers of $\alpha'/R^2$, where $R$ stands for the compactification scale, one can think of the SUGRA approximation as becoming exact when going to large volumes\footnote{Assuming that the string lift of the theory still exists.}. We shortly describe the local resolution of orbifold singularities and how to obtain the compactification spectra of them. 

\subsubsection*{Orbifold geometry}
A toroidal orbifold is constructed by choosing a six dimensional torus and modding out a subgroup of its symmetry group. So we first construct the torus from a three complex dimensional affine space by identifying points which differ by lattice vectors from a certain lattice $\Lambda_6$,
\begin{align}
 T^6 = \mathbbm{C}^3 / \Lambda_6 \,.
\end{align}
The requirement to have a $\mathbbm{Z}_7$ symmetry puts strong constraints on the lattice. In fact it has to be the $SU(7)$ root lattice with two allowed independent deformations. The $\mathbbm{Z}_7$ symmetry acts by $e_a \to e_{a+1}$ for $a=1,\ldots, 5$ and $e_6 \to - \sum_{i=1}^6 e_i$, where the $e_a$ are the simple roots. One can then combine the orbifold group with the lattice shifts to obtain the so--called space--group as a semi direct product, $S =\mathbbm{Z}_7 \rtimes \Lambda_6$. Then we can define the orbifold as 
\begin{align}
 \mathcal{O} = T^6 / \mathbbm{Z}_7 = \mathbbm{C}^3 / S \,.
\end{align}
Since the space--group action contains rotations, it will have fixed points which in the orbifold appear as curvature singularities. The fixed points  are the weights of the anti--symmetric fundamental representations, so altogether there are seven of them. Locally each of these singularities looks like $\mathbbm{C}^3/\mathbbm{Z}_7$ where the \zseven\ acts as
 \begin{align}
 \theta: (z_1,z_2,z_3) \rightarrow (\xi z_1, \xi^2 z_2, \xi^4 z_3) \qquad \text{with } \xi = e^{2\pi i /7} \,. 
 \label{eq:Z7OrbiAction}
 \end{align}
From this we see that the holomorphic three-form $\Omega = d z_1 \wedge d z_2 \wedge d z_3$ is preserved which implies $\mathcal{N}=1$ supersymmetry after compactification to four dimensions.

\subsubsection*{Heterotic strings on orbifolds}

Since the orbifold is flat everywhere except at the fixed points, a string compactification on $\mathcal O$ is described as a free conformal field theory. In order to describe a heterotic string model, one also has to embed the space--group action into the automorphism group of the left moving degrees of freedom which describe the gauge sector. In the bosonic formulation they are given by the compactification of $16$ bosons on a rigid torus $T^{16}=\mathbbm{R}^{16} / \Lambda_{16}$ with $\Lambda_{16}$ being the root lattice of $E_8 \times E_8$. Then the automorphisms act as shifts on this torus. The space--group is generated by the $\mathbbm{Z}_7$ element $\theta$ and one lattice vector $e_1$ so the embedding of $S$ into ${\rm Aut}(T^{16})$ is specified by the images of $\theta$ and $e_1$ which we call the shift vector $V$ and the discrete Wilson line $W$. From identities of the space--group we infer that $V$ and $W$ must be of order seven, i.e.\ $7V,7W \in \Lambda_{16}$. 

Now the heterotic string theory is a theory of closed strings only, so one has to sum over all boundary conditions which correspond to the conjugacy classes of the space--group $S$. These classes fall into three categories. 

First there is the identity element which corresponds to untwisted strings. The corresponding string states are the ten dimensional string states which underly certain projection conditions. The resulting four dimensional $\mathcal{N}=1$ massless spectrum contains a SUGRA sector, a super Yang--Mills sector, and chiral superfields which are the untwisted moduli, charged fields and the axion--dilaton $a^{\rm orb} - i \phi$ which plays a crucial role in the anomaly cancellation mechanism. The gauge algebra contains the full Cartan of $E_8 \times E_8$ together with the roots $P$ which satisfy $P \cdot V = P \cdot W = 0 \mmod 7$. The charges of the chiral fields are given by the winding numbers around $T^{16}$.

The second important category contains all conjugacy classes whose group action has fixed points. The corresponding massless string states turn out to be localized at the fixed points and appear as charged chiral superfields. For the $\mathbb{Z}_7$ there are 42 such classes whose representatives are $(\theta^k,(\sigma-1) e_1)$ with $k=1,\ldots,6$, $\sigma=1,\ldots,7$. However, the $(k,\sigma)$ sector always contains the CPT conjugate partners of the $(7-k,7-\sigma)$ sector so we only consider the sectors with $k=1,2,4$. From the boundary conditions we find that the charge vector of those states gets shifted by the local shift, $P_{\rm sh} = P + V_{\rm loc}$, $P \in \Lambda_{16}$, $V_{\rm loc} = k V + (\sigma-1) W$. 

The strings in third category are winding around the $T^6$ cycles so they are in general all massive and will not be considered here.

It turns out that generically in orbifold models there is one $U(1)$ gauge symmetry which appears anomalous when looking at the $4$d quantum theory. The universal axion $a^{\rm orb}$, which is the dual of the four dimensional Kalb--Ramond field, is then able to cancel this anomaly in a Green--Schwarz manner. This requires the coefficients in the anomaly polynomial to be all proportional which shows that the orbifold point is a point of high symmetry. However, the anomalous $U(1)$ has a non--vanishing Fayet--Illiopoulous term \cite{Dine:1987xk} which for a supersymmetric vacuum requires some chiral fields to attain non--trivial vacuum expectation values (vevs). Now, when twisted fields (i.e.\ those from the second category) get a vev, there is evidence that this results in a geometrical backreaction which blows up the singularity at which the field is located, resulting in a smooth but no longer flat space.

The orbifold model we are investigating here is chosen to be the one from \cite{Casas:1990yt} since it contains the standard model gauge group together with three chiral families. The shift vector and the Wilson line are
\begin{align}
 V&= \frac17 \left( 0,0,-1,-1,-1,5,-2,6 \right) \left( -1,-1,0,0,0,0,0,0 \right)\,,\nonumber \\
 W&= \frac17 \left( -1,-1,-1,-1,-1,-10,2,-9 \right) \left( 4,3,-3,0,0,0,0,0 \right)\,.
\end{align}
The non--Abelian gauge groups are $SU(3) \times SU(2) \times SO(10)$. The summary of the charged massless orbifold spectrum in terms of the non--Abelian irreducible representations (irreps) is
\begin{center}
\begin{tabular}{|r||c|c|c|c|c|c|}
 \hline
 irrep & $(\mathbf{3},\mathbf{2},\mathbf{1})$ & $(\mathbf{3},\mathbf{1},\mathbf{1})$ & $(\mathbf{\overline3},\mathbf{1},\mathbf{1})$ & $(\mathbf{1},\mathbf{2},\mathbf{1})$ & $(\mathbf{1},\mathbf{1},\mathbf{10})$ & $(\mathbf{1},\mathbf{1},\mathbf{1})$\\
 \hline
 multiplicity & 3 & 12 & 18 & 21 & 1 & 133\\
 \hline
\end{tabular}
\end{center}
\subsubsection*{Resolution of the singularities}
The resolution of orbifold singularities is a well--studied topic within algebraic geometry. In particular, toric geometry\cite{Fulton} allows to describe the resolution of a local singularity in terms of combinatorial data. For a more detailed discussion, see \cite{Lust:2006zh}. The basic idea is to add further coordinates $x_r$ together with appropriate $\mathbbm{C}^*$ scalings $\lambda_s$
\begin{align}
 (z_i,x_r) \sim (\lambda^{q_i} z_i , \lambda^{q_r} x_r) \,, \qquad \lambda^{q_i} = \prod_{s} \lambda_s^{q_i^s} \,,
\end{align}
such that the discrete orbifold action \eqref{eq:Z7OrbiAction} is induced where $x_r \neq 0$. For this case the charge assignment is
\begin{center}
 \begin{tabular}{|c|ccc|ccc|}
  \hline
  & $z_1$ &  $z_2$ &  $z_3$ &  $x_1$ &  $x_2$ &  $x_4$ \\
  \hline
  $q^1$ & $1$ & $2$ & $4$ & $-7$ & $0$ & $0$ \\ 
  \hline
  $q^2$ & $2$ & $4$ & $1$ & $0$ & $-7$ & $0$ \\ 
  \hline
  $q^3$ & $4$ & $1$ & $2$ & $0$ & $0$ & $-7$ \\
  \hline
 \end{tabular}
\end{center}
Then one removes the singular locus $\{ z_1 = z_2 = z_3 = 0 \} $ and replaces it by a set of properly intersecting hypersurfaces $E_r = \{ x_r = 0 \} $, so--called exceptional divisors, which leads to a smooth space. The geometrical orbifold is restored when one is in a region of moduli space where ${\rm Vol}(E_r)=0$. The fact that $\mathbbm{Z}_7$ is a prime orbifold implies that its blow--up topology is unique and we do not have to deal with flop transitions which can happen in non-prime orbifolds and lead to jumps in the massless spectrum \cite{Blaszczyk:2010db}. The exceptional divisors are a basis of the local homology group $H_{2,2}$ so we can use their Poincar\'e dual harmonic forms (which we also call $E_r$) to describe various $(1,1)$ forms which appear in the compactification. Then the intersection ring allows to compute topological integrals on the resolution. The basic intersection numbers are
\renewcommand{\arraystretch}{1.4}
\begin{align}
\begin{array}{r@{\;}l@{\;}r@{\;}l@{\;}}
E_1^3=E_2^3=E_4^3&=8\,,& \qquad E_1 E_2^2 = E_2 E_4^2 = E_4 E_1^2 &= 0\,, \\
E_1^2 E_2 = E_2^2 E_4 = E_4^2 E_1 &= -2 \,,& \qquad E_1 E_2 E_4 &= 1 \,.
\end{array}
\end{align}
\renewcommand{\arraystretch}{1.0}

As a next step we want to describe the resolution of the full $T^6/\mathbbm{Z}_7$ \cite{Lust:2006zh}. The global description of the resolution is rather complicated. However, since the resolution of singularities happens just locally, we can figure out the topological properties by hand. For this, we start with the orbifold and cut out small open sets around the seven fixed points. Then we replace them by the resolved local singularities which we constructed above. Therefore, we now have seven sets of three exceptional divisors, $E_{k,\sigma}$, $\sigma=1,\ldots,7$ which do not intersect when they are located at formerly different fixed points, $E_{k,\sigma} E_{l,\rho} = 0 $ if $\sigma \neq \rho$. In addition we get three inherited divisors $R_i$ which can be thought of as the duals of the forms $d z_i \wedge d \bar z_i$ from the torus which survive the orbifold projection. However, since they neither appear in the characteristic classes of the resolution nor in the expansion of the gauge flux, they are not of importance for the following discussion.

\subsubsection*{SUGRA models on the resolved space}

Since it is not known how to construct the precise metric on the resolution, we are not able to write down the string sigma model describing the compactification. Instead, we go to the low energy limit of the heterotic string, i.e.\ $\mathcal{N}=1$ SUGRA in $10$d with an \eexee\ super Yang--Mills sector, and perform the dimensional reduction. For the gauge symmetry breaking and for the appearance of chiral matter, we wrap line bundles on the resolution which are fully described by their internal field strength,
\begin{align}
 \mathcal F = H^I V^I_r E_r \,, \qquad r = (k,\sigma) \,, \quad k =1,2,4 \,, \quad \sigma = 1,\ldots,7 \,.
\end{align}
The $H^I$ are the Cartan generators of $E_8\times E_8$. The bundle vectors $V^I_r$ underly certain constraints. First of all, there are the flux quantization conditions which basically state that $7 V_r \equiv 0$ and $V_{2k,\sigma} \equiv 2 V_{k,\sigma}$ where ``$\equiv$'' means equal up to lattice vectors. Then, they have to satisfy the Bianchi identities for the Kalb--Ramond field strength,
\begin{align}
 0 = \int\limits_S\left( \tr \mathcal R^2 - \tr \mathcal F^2\right)\,, \qquad S \in \{ E_r , R_i \} \,.
\end{align}
The Bianchi identities give an upper bound on the length of the bundle vectors and fix their relative angles.

Now the topological properties of the resolution and the bundle are sufficient to compute the low energy spectrum of the compactification. The four dimensional gauge algebra is the commutant of the bundle with \eexee, i.e.\ it is spanned by the roots $P$ which are orthogonal to the bundle vectors, $P \cdot V_r = 0$. The chiral field content can be found by the Atiyah--Singer index theorem encoded in the multiplicity operator \cite{Nibbelink:2007rd},
\begin{align}
 \label{eq:multiplicityOperator}
 N =\frac16 \int\limits_X \left( \mathcal{F}^3-\frac{1}{4}\tr\mathcal{R}^2 \mathcal{F}\right)\,.
\end{align}
Acting with $N$ on the roots of \eexee\ gives the net multiplicity of the corresponding charged state. Further important contributions to the massless spectrum come from the expansion of the K\"ahler form $J$ and the Kalb--Ramond field $B_2$ in the internal $(1,1)$ forms,
\begin{align}
 J = a_i R_i-b_{r} E_{r}\,,\qquad
 B_2 =b_2 + \alpha_i R_i - \beta_{r} E_{r}\,.
 \label{eq:CmpxfdKaehler}
\end{align}
In four dimensions these fields join to form the complex scalar components of chiral multiplets, $T_i|_{\theta=0}= a_i + i \alpha_i$ and $T_r|_{\theta=0} = b_r + i \beta_r$. The real parts $a_i,b_r$ which appear in the expansion of the K\"ahler form are the K\"ahler parameters governing the size of the cycles $R_i$ and $E_r$, respectively. Furthermore, the four dimensional component $b_2$ is the dual of the universal axion $a^{\rm uni}$ in blow--up, which appears in a chiral multiplet together with the dilaton. From the gauge invariance of $H_3 = d B_2 - \Omega_{3}^{\rm YM} + \Omega_{3}^{\rm L}$, we see that the model dependent axions $\beta_{r}$ have to transform with a shift as $\delta \beta_{r} = V^I_{r} \chi^I$, where $\chi^I$ are the gauge parameters. For the $\alpha_i$ one finds that they do not transform.

In a blow--up model generically many of the $U(1)$'s are broken which can be understood in two ways. When one comes from the orbifold point and assigns vevs to charged twisted states this leads to a gauge symmetry breaking via the Higgs effect. On the pure blow--up side one identifies these $U(1)$'s as the structure group of the Abelian bundle so classically they would still be intact. However, it turns out that the chiral spectrum renders them anomalous. Then the many model dependent axions $\beta_r$ are able to cancel all these anomalies due to their shift transformation, which comes at the cost of a St\"uckelberg--like mass term for the gauge bosons. This is the same mechanism which is at work on the orbifold to cancel the one anomalous $U(1)_A$ using the orbifold axion $a^{\rm orb}$. In this light, the gauge group which is usually referred to as ``anomalous'' $U(1)$ should best be called broken $U(1)$ with canceled anomaly, as precisely due to the Green--Schwarz mechanism all anomalies are canceled (on the orbifold and in blow--up).

The bundle vectors which specify the resolution model we are working with are given as blow--up mode charges at the end of appendix \ref{sec:AppendixSpectrumComparison}.
The non--Abelian gauge algebra in blow--up is $SU(3) \times SU(2)\times SO(10)$. A short summary of the charged spectrum is
\begin{center}
\begin{tabular}{|r||c|c|c|c|c|c|}
 \hline
 irrep & $(\mathbf{3},\mathbf{2},\mathbf{1})$ & $(\mathbf{3},\mathbf{1},\mathbf{1})$ & $(\mathbf{\overline3},\mathbf{1},\mathbf{1})$ & $(\mathbf{1},\mathbf{2},\mathbf{1})$ & $(\mathbf{1},\mathbf{1},\mathbf{10})$ & $(\mathbf{1},\mathbf{1},\mathbf{1})$\\
 \hline
 multiplicity & 3 & 10 & 16 & 17 & 1 & 86\\
 \hline
\end{tabular}
\end{center}

\section{Spectrum matching}
\label{sec:SpectrumMatching}
In this section we want to compare the spectra on the orbifold and in blow--up. We will see how to uncover all orbifold states in blow--up after a suitable field redefinition. Some of the orbifold states, which couple to the blow--up modes that get a vev in the blow--up procedure, acquire a mass via the Higgs mechanism and are thus removed from the massless spectrum. By evaluating the Atiyah--Singer index theorem locally at each (compact) exceptional divisor, we can calculate the particle spectrum at each of the seven different fixed points separately (but not separately for the twisted sectors). In this way we can identify vector--like pairs of states if they reside at different fixed points. These states cannot be seen by using the index theorem on the entire Calabi--Yau geometry. The reason that we can trust the index theorem locally is that the Bianchi identities are satisfied locally, because there are only localized exceptional divisors in the resolution of the prime orbifolds. 
\subsection{Field redefinitions}
When comparing the spectrum on the orbifold and in blow--up, one faces the problem that the states on the orbifold $\Phi_{\gamma}^\text{Orb}$ are characterized via their shifted momenta, while the states in blow--up $\Phi_{\gamma}^\text{BU}$ are characterized via \eexee\ lattice vectors ($\gamma=(k,\sigma,i)$ labels all a priori massless states $i$ on the orbifold at all fixed points $(k,\sigma)$). In the following, we distinguish between fields on the orbifold which generate the blow--up by attaining non--zero vevs and fields on the orbifold which stay free fields in blow--up. We refer to the former fields as blow--up modes. From the transformation behavior of the localized axions we can identify them with the complexified K\"ahler moduli as
\begin{align}
 \Phi_{r}^{\text{BU-Mode}}=e^{b_r+i\beta_r}\,,
\end{align}
where $b_r$ are the K\"ahler moduli parameterizing the size of the blown up cycle and $\beta_r$ are the model dependent axions, cf.\ \eqref{eq:CmpxfdKaehler}. A connection to the latter fields is established by considering field--redefinitions via exponentiating
\begin{align}
 \Phi_{\gamma}^\text{BU}=e^{-\sum_k r_{k,\sigma}^\gamma (b_{k,\sigma}+i\beta_{k,\sigma})}\Phi_{\gamma}^\text{Orb}\,,
 \label{eq:FieldRedef}
\end{align}
where the coefficients $r^\gamma_{k,\sigma}$ appearing in the linear combination are specified below in \eqref{eq:FieldRedefs}. The sum over the twisted sectors $k=1,2,4$ in the redefinition allows for the occurrence of twisted fields which live at the same fixed point $\sigma$ but in different twisted sectors of the orbifold theory. Our conventions relating the orbifold charges to the blow--up charges after the field redefinition are the following: We denote the charges under the 16 Cartan generators of the orbifold states which become blow--up modes by $q_I^{k,\sigma}$, $I=1,\ldots,16$. They coincide with the shifted momenta of the orbifold states and by construction with the line bundle vectors $V_{k,\sigma}^I$. The charges of the other fields on the orbifold are denoted by $Q_I^\gamma$ and the redefined charges in blow--up by $Q'^\gamma_I$. The difference between the charges $Q_I^\gamma$ and $Q'^\gamma_I$ is $\Delta_I^\gamma$,
\begin{align}
\label{redef}
Q'^{\gamma}_I=Q^{\gamma}_I-\Delta^{\gamma}_I,\qquad \Delta^{\gamma}_I=\sum\limits_{k=1,2,4} r_{k,\sigma}^\gamma q^{k,\sigma}_I\,.
\end{align}
\indent The exponential in \eqref{eq:FieldRedef} leads to the correct behavior under gauge transformations: as the axions $\beta_{k,\sigma}$ transform with a shift, the redefined states transform linearly
\begin{align}
 \beta_{k,\sigma}\rightarrow\beta_{k,\sigma}+V_{k,\sigma}^I\chi_I \qquad \Rightarrow \qquad \Phi_{\gamma}^\text{BU}\rightarrow e^{i\chi_I (Q_I^\gamma-\Delta_I^\gamma)} \Phi_{\gamma}^\text{BU}\,,
 \label{eq:FieldRedefTrafo}
\end{align}
where $\chi_I$ is the gauge parameter. The exponential map also has the surprising effect that the blow--down limit is recovered by taking the K\"ahler parameters $b_{k,\sigma}$ governing the size of the exceptional cycles to $-\infty$ rather than to $0$. The more intuitive behavior of $b_{k,\sigma}\rightarrow 0$ in blow--down can be obtained by constructing a different measure for the volume of curves \cite{Aspinwall:1993xz}.

\indent Let us have a closer look at the field redefinitions \eqref{eq:FieldRedef}. An arbitrary combination of twisted sectors does not lead to a consistent field redefinition. When matching the states on the orbifold with those in blow--up we compare the gauge charges of the states on both sides. From the transformation property \eqref{eq:FieldRedefTrafo} it is apparent that the difference between the orbifold shifted momentum and the line bundle vector precisely corresponds to the gauge charges. Since the spectrum computation is carried out by evaluating an index theorem for the $480$ root vectors of the theory, only those redefinitions for which the difference yields a \eexee\ root vector correspond to a proper redefinition. We find that the following possible charge redefinitions are realized:
\begin{subequations}
 \label{eq:FieldRedefs}
 \begin{align}
  &Q_{k,\sigma}^{\text{Orb}}\mapsto Q_{k,\sigma}^{\text{BU}}=Q_{k,\sigma}^{\text{Orb}}-V_{k,\sigma}\,,\label{eq:redef1}\\[3mm]
  &Q_{k,\sigma}^{\text{Orb}}\mapsto Q_{k,\sigma}^{\text{BU}}=Q_{k,\sigma}^{\text{Orb}}+V_{l,\sigma}+V_{m,\sigma}\,, \quad k \neq l \neq m \neq k\,,\label{eq:redef2}\\[3mm]
  &\!\!\!\begin{array}{l}
   Q_{1,\sigma}^{\text{Orb}}\mapsto Q_{1,\sigma}^{\text{BU}}=Q_{1,\sigma}^{\text{Orb}}+V_{1,\sigma}-V_{2,\sigma}\,,\\[2mm]
   Q_{2,\sigma}^{\text{Orb}}\mapsto Q_{2,\sigma}^{\text{BU}}=Q_{2,\sigma}^{\text{Orb}}+V_{2,\sigma}-V_{4,\sigma}\,,\\[2mm]
   Q_{4,\sigma}^{\text{Orb}}\mapsto Q_{4,\sigma}^{\text{BU}}=Q_{4,\sigma}^{\text{Orb}}-V_{1,\sigma}+V_{4,\sigma}\,,
  \end{array}\label{eq:redef3}\\[3mm]
  &\!\!\!\begin{array}{l}
   Q_{1,\sigma}^{\text{Orb}}\mapsto Q_{1,\sigma}^{\text{BU}}=Q_{1,\sigma}^{\text{Orb}}+V_{1,\sigma}+V_{2,\sigma}-V_{4,\sigma}\,,\\[2mm]
   Q_{2,\sigma}^{\text{Orb}}\mapsto Q_{2,\sigma}^{\text{BU}}=Q_{2,\sigma}^{\text{Orb}}-V_{1,\sigma}+V_{2,\sigma}+V_{4,\sigma}\,,\\[2mm]
   Q_{4,\sigma}^{\text{Orb}}\mapsto Q_{4,\sigma}^{\text{BU}}=Q_{4,\sigma}^{\text{Orb}}+V_{1,\sigma}-V_{2,\sigma}+V_{4,\sigma}\,.\label{eq:redef4}
  \end{array}
 \end{align}
\end{subequations}
\subsection{Local massless particle spectrum}
Now we turn to the calculation of the massless particle spectrum. On the orbifold this is done by explicitly constructing all shifted momenta which fulfill the masslessness and level matching condition. Finding the massless particle spectrum in blow--up is harder, but can be done with the help of the Atiyah--Singer index theorem \eqref{eq:multiplicityOperator}.\\
\indent Due to the compactness of the exceptional divisors which support the gauge flux $\mathcal{F}$, the expression for the index \eqref{eq:multiplicityOperator} is simply the sum of the contributions at the seven fixed points. Knowledge of all intersection numbers from toric geometry allows us to evaluate the integral explicitly. At each fixed point $\sigma$ the index can be written as
\begin{align}
 \label{eq:multiplicityOperatorExplicit}
 N(\sigma) = \frac{1}{3} \sum\limits_{k=1,2,4} [4 H_{k,\sigma}^3 - H_{k,\sigma}] - H_{1,\sigma} H_{2,\sigma}^2 - H_{1,\sigma}^2 H_{4,\sigma} - H_{2,\sigma} H_{4,\sigma}^2 + H_{1,\alpha} H_{2,\sigma} H_{4,\sigma}\,,
\end{align}
where we used the short--hand notation $H_{k,\sigma}=V_{k,\sigma}^{I} H_I$ with line bundle vectors $V_{k,\sigma}^{I}$ and Cartan generators $H_I$. The overall index is obtained by summing the above expression over all $7$ fixed points, $N=\sum_\sigma N(\sigma)$. It is obvious that while $N(\sigma)$ can be evaluated for every fixed point $\sigma$ separately, the expression still contains a sum over the twisted sectors $k$, hence the index theorem is blind to the twisted sector to which the state originally belonged. To determine the multiplicity of all states in blow--up one acts with $N$ on all \eexee\ roots. For this reason we also refer to \eqref{eq:multiplicityOperatorExplicit} as (local) multiplicity operator.\\
\indent In order to calculate the particle spectrum, we can thus proceed as follows. First we check under which irrep of the unbroken gauge groups in blow--up the 480 root vectors of the theory transform. Then we act with $N$ on the roots. This yields the multiplicity of each massless SUSY matter multiplet in blow--up. As \eqref{eq:multiplicityOperatorExplicit} is an odd polynomial in $H_I$, the multiplicity changes sign for CPT conjugate states. In the rest of the paper, we evaluate multiplicities by acting on the highest weight of fundamental representations and on the lowest weight of the CPT conjugate of anti--fundamental representations. Hence states transforming in fundamental representations are assigned positive multiplicity and states transforming in anti--fundamental representations are assigned negative multiplicity.
\subsection{Spectrum comparison}
As described in the previous subsection, it should in principle be possible to match all orbifold states to blow--up states via field redefinitions of the type \eqref{eq:FieldRedefs}. The details of matching the particle spectra on both sides are worked out in this subsection. A table of all \eexee\ root vectors, their redefinition, and the corresponding orbifold states can be found in appendix \ref{sec:AppendixSpectrumComparison}.\\
\indent Unfortunately, there are several obstacles that have to be overcome. First, determining the particle spectrum in blow--up using the index theorem is not as powerful as computing it directly from the shift vectors on the orbifold. The reason is that the index theorem can only count the net number of left--chiral minus right--chiral states while the explicit orbifold calculation reveals states of both chiralities separately. Second, the blow--up is generated by assigning non--trivial vevs to twisted orbifold states. This leads to effects familiar from the Higgs mechanism: the gauge groups under which the fields that get a vev are charged get broken and the vevs of the fields provide a mass term for other fields that couple to the higgsed field.\\
\indent The problem that the rank of the non--Abelian gauge group is reduced in blow--up via the Higgs mechanism is avoided by choosing only non--Abelian gauge singlets as blow--up modes. If this were not possible, either because there is no solution to the Bianchi identities which involve only twisted singlets as blow--up modes or because there are fixed points without twisted singlet matter, one would have to reconstruct the breaking of the non--Abelian gauge groups by group--theoretical means. While the matching is still possible also in this case, we refrained from doing so in the example in order to keep the exposition as simple as possible.\\
\indent For solving the problem of the vector--like states that are not captured by the index theorem, the local multiplicity operator is of huge importance. It happens quite often that several different orbifold states are redefined via \eqref{eq:FieldRedefs} to the same root vector, while other roots do not occur at all in the redefinition process. The latter manifests itself by yielding a multiplicity of zero when one acts on such a root with the multiplicity operator. The former leads to a multiplicity which is in general not equal to one. If there are states which are redefined to the same root, while others are redefined to the negative root (i.e.\ the charge conjugate one), the multiplicity operator will only see the number of the one states minus the number of the other states so we do not see vector--like pairs. This leads to the effect that there are seemingly less states in blow--up than on the orbifold. The big advantage of the local multiplicity operator is now that even these vector--like pairs can be identified as long as they do not in addition live at the same fixed point on the orbifold. Additionally, by checking their dependence on the K\"ahler parameters $b_{r}$, it can be checked which states are expected to get a mass in blow--up. By direct inspection of the Yukawa couplings on the orbifold side we then verify that all involved states couple to one or more fields that act as blow--up modes. Computation of the anomalies on both sides of the theory in the next section provides a very strong cross--check that the identified mass terms are indeed correct. Incidentally, the motivation for matching all states between the orbifold and the blow--up theory was driven by anomaly considerations.
\subsubsection*{Matching of massless states}
\indent In order to illustrate the methods for matching the spectra explained above, let us look at examples from the table of appendix \ref{sec:AppendixSpectrumComparison}. Let us begin with the 3 quark doublets $(\mathbf{3},\mathbf{2},\mathbf{1})$. The first field $Q_1$ lives in the untwisted sector. Hence it does not need to be redefined. The local multiplicity operator tells us that it lives to $1/7$ at each of the 7 fixed points, i.e.\ the field is democratically smeared out over all fixed points, as one would expect for an untwisted field. The fields $Q_2$ and $Q_3$ both live at the first fixed point. Both are redefined to a unique root vector via \eqref{eq:redef1} at the first fixed point (and of the second respectively first twisted sector). By looking at the local multiplicity operator, we see a multiplicity of one at the first fixed point. Hence the local multiplicity operator exactly sees the orbifold state. At the other fixed points, we see fractional multiplicities of $\pm 1/7$, which however sum to zero and thus the overall multiplicity is one. These non--existing states can be interpreted as those untwisted states which were projected out on the orbifold.
As long as they sum to zero, we will ignore them in the following. If they do not sum to zero but to one, they indicate an untwisted sector field, as seen for the field $Q_1$.\\
\indent For the triplets $(\mathbf{3},\mathbf{1},\mathbf{1})$ there are states that transform in the fundamental $\mathbf{3}$ as well as in the anti--fundamental $\mathbf{\overline{3}}$. 
We conventionally only look at the triplet weights since the anti--triplets weights correspond to their negatives. Thus, a positive multiplicity indicates a triplet state whereas a negative multiplicity stands for the presence of an anti--triplet state.
An example for this are e.g. the states $\overline{t}_7$ and $t_6$ which transform in the $(\mathbf{\bar{3}},\mathbf{1},\mathbf{1})$ and $(\mathbf{3},\mathbf{1},\mathbf{1})$. Their overall multiplicity is -1 and 1, and the local multiplicity operator reveals that these states live at fixed points 7 and 6, respectively.\\
\indent Something conceptually new happens for the orbifold states $t_5$, $t_{12}$, $\overline{t}_{11}$, and $\overline{t}_{18}$. Albeit these four states are redefined to the same root the total multiplicity is zero. This happens because the multiplicity operator can only count the net number of states which is $2-2=0$. However, the local multiplicity operator gives some insight into what is happening. The three states $t_5$, $t_{12}$, $\overline{t}_{11}$ all live at fixed point 5 on the orbifold. As there are two left--chiral and one right--chiral state the local multiplicity is 1. For the one right--chiral state $\overline{t}_{18}$, there is a local multiplicity of -1 at fixed point 6. Hence the overall multiplicity is zero. The multiplicities of the other states can be worked out in a similar manner.
\subsubsection*{Matching of massive states}
Vector--like states can acquire a mass in the blow--up procedure from trilinear Yukawa couplings. The selection rules for allowed Yukawa couplings on the orbifold arise from requiring gauge invariance, compatibility with the space--group, and conservation of H--momentum. Conservation of $R$--charge will be discussed below. Gauge invariance simply amounts to the requirement that the sum of the left--moving shifted momenta of the strings involved in the coupling is zero.\\
\indent The space--group selection rule amounts to the requirement that the product of the constructing space--group elements of the states involved in the Yukawa coupling must be the identity element $(\mathbbm{1},0)$. For trilinear couplings this states that the allowed couplings are of the form
\begin{align}
 (k=1,\sigma_1) \circ (k=2,\sigma_2) \circ (k=4,\sigma_4) \,, \qquad \text{with} \ \sigma_1+ 2\sigma_2 + 4\sigma_4  =0\mmod 7\,.
\label{spacegroup}
\end{align}
If the coupling involves states which reside all at the same fixed point ($\sigma_1=\sigma_2=\sigma_4$), the space--group selection rule is trivially fulfilled. However, there also exist solutions to \eqref{spacegroup} for states coming from three different fixed points. Since these couplings arise from world--sheet instantons \cite{Dixon:1986qv,Hamidi:1986vh}, they are suppressed by a factor of the form $e^{-a_i}$ where $a_i$ are the moduli which govern the sizes of the orbifold or Calabi--Yau (cf.\ \eqref{eq:CmpxfdKaehler}). As it turns out, in our case H--momentum is conserved for the trilinear couplings if the space--group selection rule is fulfilled.\\
\indent There can actually be more selection rules coming from the internal part of the Lorentz group. For a local orbifold $\mathbbm{C}^3/\mathbbm{Z}_N$ the rotation of the three individual complex planes is a continuous symmetry. Since the invariant spinor is charged under it, this symmetry will be an $R$--symmetry. The charges are computed as
\begin{align}
 R_\gamma^i=q_{\text{sh},\gamma}^i + N_\gamma^i - \overline{N}_\gamma^i\,,
\label{eq:RCharge}
\end{align}
where $q_\text{sh}$ are the shifted right--moving internal momenta of the orbifold state $\Phi_\gamma^\text{Orb}$ and $N$ ($\overline N$) are the (anti--) holomorphic oscillator numbers. The conservation rule reads
\begin{align}
 \sum_\zeta R_\zeta^i = 1 \,,
\label{eq:RChargeRule}
\end{align} 
where $\zeta$ runs over the three states involved in the Yukawa coupling. Equation \eqref{eq:RChargeRule} is trivially fulfilled for states without oscillators if the space--group rules are. However, in a compact orbifold this symmetry will be broken down to a subgroup by the torus lattice. Therefore the formerly forbidden couplings are expected to be surpressed by the size of the lattice. If the lattice is factorizable, the remaining symmetry is the discrete rotation of the three two--tori. In this case the selection rule needs only be satisfied up to multiples of the order of the orbifold group. For the non--factorizable $SU(7)$ lattice of the \zseven\ orbifold, we checked that the symmetry is broken completely except for the \zseven\ itself, so \eqref{eq:RChargeRule} should not be imposed on the orbifold.\\
\indent The SUGRA theory on the blow--up side is, however, only valid in the large volume limit. In particular, we expect that the $R$--charge selection rule \eqref{eq:RChargeRule}, which is broken by the orbifold lattice, is still a valid symmetry in the large volume limit. Therefore we expect the states, which are supposed to get a mass via such suppressed couplings on the orbifold, to appear as massless states in the multiplicity operator in blow--up. By comparing the spectra we indeed find that the index theorem sees massless states for which the orbifold theory predicts non--local mass terms or mass terms which do not satisfy \eqref{eq:RChargeRule}. To illustrate the absence of both types of mass terms in blow--up we look at suitable examples.\\
\indent As an example for mass terms not satisfying \eqref{eq:RChargeRule} consider the singlet states $s_{25}$, $s_{26}$, $s_{70}$, $s_{111}$, $s_{112}$ and $s_{113}$, see appendix \ref{sec:AppendixSpectrumComparison}. These states are all oscillator states which explains their degeneracy and which makes them sensible to a possible $R$--symmetry. Together with the blow--up modes $s_{68}$ and $s_{27}$, there are the following orbifold trilinear superpotential couplings when imposing only gauge-- and space--group invariance and the H--momentum rule:
\begin{align}
 \left( s_{111} \ s_{112} \ s_{113} \right) \begin{pmatrix} a_{11} s_{68} & a_{12}s_{68} & a_{13}s_{27} \\  a_{21}s_{68} & a_{22}s_{68} & a_{23}s_{27} \\  a_{31}s_{68} & a_{32}s_{68} & a_{33}s_{27} \\ \end{pmatrix} \begin{pmatrix} s_{25} \\  s_{26}\\ s_{70} \end{pmatrix}\,,
\end{align}
where the $a_{ij}$ are coefficients of order one. Now when one gives a vev to the blow--up modes $s_{68}$ and $s_{27}$, these couplings become a rank three mass matrix and thus one would expect all 6 singlets to become massive and disappear from the chiral spectrum in blow--up. However, when we look at the roots to which these singlets can be redefined, the local multiplicity operator reveals that there are four states at the resolved fixed point where the singlets in question were localized. Therefore four of these singlets must stay massless during blow--up. This means that the above mass matrix must only have rank one, such that just one pair of singlets is decoupled. One could explain this by assuming that all coefficients $a_{ij}$ are equal, but this assumption is a priori not justified and would lead to mixing of the fields during redefinition. The correct explanation is to argue that the local multiplicity operator sees states only in the large volume limit where the $R$--symmetry \eqref{eq:RChargeRule} is exact. Imposing $R$--symmetry here would set all coefficients to zero except for $a_{21}$ and $a_{23}$ and therefore naturally explain the rank one mass matrix at this place.\\
\indent To illustrate the non--local mass terms, we investigate the triplet states $t_5$, $t_{12}$, $\overline{t}_{11}$, and $\overline{t}_{18}$ encountered above. From the employed redefinitions we find
\begin{subequations}
\label{eq:tripletMassTerms}
 \begin{alignat}{3}
 t_5^{\,\rm BU} \overline t_{11}^{\,\rm BU} &= t_5^{\,\rm Orb} \overline{t}_{11}^{\,\rm Orb} e^{-b_{4,5}+b_{1,5}+b_{4,5}} &&= t_5^{\,\rm Orb} \overline{t}^{\,\rm Orb}_{11}e^{b_{1,5}}\,,\label{eq:tripletMassTerm1}\\
  t_{12}^{\,\rm BU} \overline t_{11}^{\,\rm BU} &= t_{12}^{\,\rm Orb} \overline{t}_{11}^{\,\rm Orb} e^{-b_{1,5}+b_{1,5}+b_{4,5}} &&= t^{\,\rm Orb}_{12} \overline{t}_{11}^{\,\rm Orb}e^{b_{4,5}}\,.\label{eq:tripletMassTerm2}
 \end{alignat}
\end{subequations}
The coupling of $t_5$ and $t_{12}$ with $\overline{t}_{18}$ is non--local as the states reside at different fixed points. Hence this coupling is not captured by the multiplicity operator. The redefinitions clearly show that in blow--up where $b_{k,\sigma}\rightarrow\infty$, the couplings \eqref{eq:tripletMassTerms} provide a mass term which vanishes in the blow--down limit $b_{k,\sigma}\rightarrow-\infty$. This means that from the blow--up perspective a linear combination of $t_5$ and $t_{12}$ pairs up with $\overline{t}_{11}$ and lifts the exotic state from the massless particle spectrum in blow--up. This behavior is also  confirmed from the orbifold perspective. The appearance of $b_{1,5}$ \eqref{eq:tripletMassTerm1} shows that $t_5$ from the $\theta^4$ sector and $\overline{t}_{11}$ from the $\theta^2$ sector couple to the blow--up mode from the $\theta$ sector as dictated by the space--group selection rule. Likewise, for the second mass term \eqref{eq:tripletMassTerm2} we find a coupling between  $t_{12}$ from the $\theta$ sector, $\overline{t}_{11}$ from the $\theta^2$ sector, and the blow--up mode from the $\theta^4$ sector as indicated by $b_{4,5}$.\\
\indent The local $R$--charge selection rule \eqref{eq:RChargeRule} is only relevant for oscillator states, as states satisfying the space--group selection rule have $\sum_\zeta q_{\text{sh},\zeta}^i=1$ and hence \eqref{eq:RChargeRule} is fulfilled for states without oscillators. Interestingly, the states which have oscillators often allow for more than one possible redefinition \eqref{eq:FieldRedefs}. Imposing \eqref{eq:RChargeRule} in conjunction with consistency of the local blow--up spectra singles out a unique field redefinition. Using these redefinitions, we were finally able to establish a perfect match between the anomalies on the orbifold and in blow--up, which we take as a strong cross--check that the above discussion is valid. This will be explained in the next chapter.\\
\indent The above analysis has been carried out in a similar fashion for all other $\mathcal{O}(200)$ states. Each time we find mass terms of the form \eqref{eq:tripletMassTerms} from the redefinitions on the blow--up side, they also constitute allowed couplings on the orbifold side and lead in the end to a perfect match of the anomaly computation. We expect also that there exists an orbifold mechanism explaining why a local $R$--charge can be applied in this case. This is still work in progress and will be discussed in the future.

\section{Anomalies}
\label{sec:Anomalies}
As explained in section \ref{sec:SpectrumMatching} the difference between the spectrum on the orbifold and in blow--up 
can be understood through field redefinitions which involve the blow--up modes. 
The change of the spectrum away from the orbifold point can also be investigated by studying the change of the anomalies \cite{Nibbelink:2007ew, Nibbelink:2008tv}. We consider the anomaly cancellation mechanism of the four dimensional effective theory. This mechanism is understood in terms of the universal and the non--universal (localized) axions. On the orbifold there is a single anomalous $U(1)$ and a single axion to cancel it. But the displacement from the orbifold point through singlet vevs generically causes all the other $U(1)$s to become anomalous as well. Also, the change of the massless spectrum in blow--up reflects in the change of the anomaly. Field redefinitions help us to understand how the original anomaly polynomial changes and how the Green--Schwarz mechanism occurs via the blow--up modes, which are interpreted as localized axions. The anomaly universality condition on the orbifold \cite{Casas:1987us} reflects the fact that the $U(1)$--gravitational anomalies $U(1)\times\text{grav}^2$, the mixed $U(1)$--non--Abelian anomalies $U(1)\times G^2$ (with arbitrary non--Abelian gauge group $G$), and the pure $U(1)$ anomalies $U(1)_I\times U(1)_J\times U(1)_K$ are canceled with a single axion $a^{\text{orb}}$. As explained in the previous sections, in blow--up we expect to find, besides the universal axion $a^\text{uni}$, non--universal, localized axions which contribute to the cancellation.

For analyzing anomalies in the 4d effective field theory we study the 4d anomaly polynomial $I_6$ and the 4d Green--Schwarz mechanism \cite{Green:1984sg} derived from the one of ten dimensional supergravity. The variation of the $B_2$ field leaves $H_3$ invariant and causes the counter--term $\delta S_B=\delta \int B_2 X_8$ to cancel the anomalous variation of the action. The ten dimensional anomaly $G=\int I_{10}$ for the 10d gauginos is encoded in a 12--form $I_{12}$ through the descent equations $dI_{10}=\delta I_{11},\ d I_{11}=I_{12}$\cite{Green:1984sg}. Dimensional reduction of $I_{12}$ and $S_B$ leads to the 4d anomaly cancellation in terms of the various axions \cite{Nibbelink:2007ew} that descend from the $B_2$ expansion \eqref{eq:CmpxfdKaehler}. The form $I_{12}$ is factorized in terms of a 4--form $X_4$ and an 8--form $X_8$ as 
\begin{align}
I_{12}&=X_4X_8\,,\nonumber\\
 X_8&=\frac{1}{4}\left((\text{tr} \mathfrak{F}'^2)^2+(\text{tr} \mathfrak{F}''^2)^2\right)-\frac{1}{4}\text{tr}\mathfrak{F}'^2\text{tr}\mathfrak{F}''^2-\frac{1}{8}(\text{tr}\mathfrak{F}'^2+\text{tr}\mathfrak{F}''^2)\tr \mathfrak{R}^2
+\frac{1}{8}\tr \mathfrak{R}^4+\frac{1}{32}(\tr \mathfrak{R}^2)^2\,,\label{I12}\\
X_4&=\tr \mathfrak{R}^2-\tr\mathfrak{F}'^2-\tr\mathfrak{F}''^2\,.\nonumber
\end{align}
We denote the 10d spin connection with $\mathfrak{W}$, 10d curvature with $\mathfrak{R}$, 10d gauge fields with $\mathfrak{A}$, and 10d gauge field strengths with $\mathfrak{F}$. When it is necessary to distinguish between the two \ee s, we mark the gauge fields and the field strengths in the first and second \ee\ with $'$ and $''$, respectively. The 10d quantities are decomposed in terms of 6d internal and 4d components as $\mathfrak{W}=\mathcal{W}+\omega,\ \mathfrak{R}=\mathcal{R}+R, \ \mathfrak{A}=\mathcal{A}+A,\ \mathfrak{F}=\mathcal{F}+F$, where the first term and the second term in the sums are the 6d and 4d components, respectively. Starting from the anomaly \eqref{I12} in ten dimensions, our aim is to understand the anomaly cancellation mechanism in a compactification away from the \zseven\ orbifold point. 

\subsection{Non--universal axions}
\label{subsec:nonUniversalAxions}

Here we would like to show explicitly how the cancellation is implemented once the blow--up is performed. We show the way in which axions arise when Abelian gauge bundles are present in the internal manifold. This has been studied in \cite{Nibbelink:2007ew} for a non--compact resolution with a single non--universal axion and in \cite{Blumenhagen:2005ga} in a more generic formulation. Our analysis applies these results to a case where multiple non--universal axions appear in a blow--up of an MSSM--like model on a compact orbifold.

The change of the effective action due to gauge transformations (parameterized by $\chi$) and Lorentz transformations (parameterized by $\Theta$) is given by \cite{GSW2}
\begin{align}
\label{eq:effActionChange}
\ G(\chi,\Theta)=\int \limits_{\mathcal{M}\times M^4} I_{10}=\int \limits_{\mathcal{M}\times M^4} \left(\tr(\Theta d \mathfrak{W})-\tr(\chi d \mathfrak{A}) \right)X_8\,.
\end{align}
where we split up the 10d space into the 6d internal manifold $\mathcal{M}$ and 4d Minkowski space $M^4$, and omit a numerical factor arising in the dimensional reduction. The variation of the axion field $\delta_{\chi,\Theta} B_2=-\tr(\Theta d \mathfrak{W})+\tr(\chi d\mathfrak{A})$ induces a variation $\delta S_B$ which exactly cancels $G(\chi,\Theta)$. In the compactification to 4d, the anomaly cancellation arises from the variations of the $B_2$ components inherited from 10d variations, and from imposing the condition $\delta_{\chi_{0}} d B_2=0$, where $\chi_{0}$ are gauge transformations on the gauge bundle $\mathcal{A}\rightarrow \mathcal{A}+\delta_{\chi_{0}}\mathcal{A}, \ [\mathcal{A},\chi_{0}]=0$. The $B_2$ field is expanded as in \eqref{eq:CmpxfdKaehler}.

The 4d \textit{universal} axion $a^\text{uni}$ and the \textit{non--universal} axions $\beta_r$ cancel the 4d anomaly. This can be seen from the reduction of $G$ and $S_B$ and by performing a field redefinition necessary to ensure $\delta_{\chi_{0}} d B_2=0$. The dimensional reduction of the variation of the effective action \eqref{eq:effActionChange} reads
\begin{align}
I_4&=\int\limits_{\mathcal{M}} \tr (\Theta d\mathfrak{W}) X_8-\tr(\chi dA)\int\limits_{\mathcal{M}}X_{6,2}- \int\limits_{\mathcal{M}}\tr(\chi \mathcal{F})X_{4,4}\,,\\
G&=\int\limits_{M^4} I_4=
\int\limits_{M^4} \left[\tr (\Theta d\omega) X^{\rm uni}_2+\Theta_a(\mathcal{W}_a^rX^r_4+\mathcal{W}^i_aX^i_4)\right]-\int\limits_{M^4}\left[\tr(\chi dA)X^{\rm uni}_2+V^I_r \chi_I X_4^r\right]\,.
\end{align}
The forms $X_{2k,2l}$ with $2(k+l)=8$ are the sum of all the terms in $X_8$ having $2k$ indices in the internal space, and $2l$ indices in the external 4d space, and $X_4^r=\int_{\mathcal{M}} E_r X_{4,4}$. Furthermore, $\Theta=\Theta_a T^a$ is the expansion of the Lorentz transformation in terms of $SO(9,1)$ generators $T^a$ and $d\mathcal{W}=(\mathcal{W}^r_a E_r+\mathcal{W}^i_a R_i)T^a$ is the expansion of the derivative of the spin connection in $T^a$ and in (1,1) forms on the internal manifold.\\
\indent The whole anomaly variation of the action can be divided into a \textit{universal} and a \textit{non--universal} part given by
\begin{align}
G_\text{uni}&=\int\limits_{M^4} \left( \tr(\Theta d\omega)-\tr(\chi dA)\right)X^{\rm uni}_2\,,\\
G_\text{non}&=\int\limits_{M^4}\Theta_a(\mathcal{W}^r_a X^r_4+\mathcal{W}^i_a X^i_4)-V^I_r \int\limits_{M^4} \chi_I X^r_4\,,
\end{align}
where $X^i_4=\int\limits_{\mathcal{M}}R_i X_{4,4}$. Along the same lines one can write the dimensional reduction of $S_B$ as 
\begin{align}
 S_B=\!\!\int\limits_{M^4\times\mathcal{M}}\!\! B_2 X_8=\!\!\int\limits_{M^4\times\mathcal{M}}\!\! b_2 X_{6,2}+\!\!\int\limits_{M^4\times\mathcal{M}}\!\!(\alpha_iR_i+\beta_r E_r)X_{4,4}=\!\!\int\limits_{M^4}\!\! b_2 X_2^{\rm uni}+\!\!\int\limits_{M^4}\!\!( \beta_rX^r_4+\alpha_iX^i_4)\,.
\end{align}

Now we can understand how the 4d transformations of $a^\text{uni},~\beta_r$, and $\alpha_i$ inherit the 10d anomalous variations of the $B_2$ field. Considering the 4d variations of the axions to be exactly the same as those of $B_2$, and without taking into account mixed index variations (which is equivalent to a redefinition of $B_2$ in order to achieve $\delta_{\chi_{0}}B_2=0$), anomaly cancellation in 4d is implemented by
\begin{align}
\delta b_2&=\tr(\chi dA)-\tr(\Theta d\omega)\,,\\
\delta B_{1,1}&=\tr(\chi \mathcal{F})-\tr(\Theta d\mathcal{W})= \chi^IV^I_rE_r-\Theta_a (\mathcal{W}^r_a E_r-\mathcal{W}^i_a R_i)\,,
\end{align}
where $B_{1,1}= \alpha_i R_i+\beta_r E_r$. The $\alpha_i$ and $\beta_r$ satisfy
\begin{align}
\delta \alpha_i=-\Theta_a \mathcal{W}^i_a,\ \ \delta \beta_r=\chi^I V^I_r+\Theta_a \mathcal{W}^r_a\,,
\end{align}
which ensures 
\begin{align} 
G_\text{non}+G_\text{uni}+\delta_{\chi} \int\limits_{M^4\times\mathcal{M}} B_2 X_8=0\,.
\end{align}
\indent Let us now take a complementary approach, which proceeds via studying the reduction of $H_3$ and checking how $\delta H_3$ is canceled by the variation of the 4d axions. Let us consider gauge variations only. This will clarify why it is allowed to restrict the variation of $B_2$ to the 4d axions $\beta_r$ or $a^\text{uni}$.\\
\indent The three--form $\Omega_3^{\text{YM}}=\tr(\mathfrak{A}\mathfrak{F}-\mathfrak{A}^3/3)$ can be expanded in terms of 4d and 6d parts as
\begin{align}
\Omega_3^{\text{YM}}= \Omega_3^{\text{YM},\text{4d}}+\tr(\mathcal{A}d \mathfrak{A})+\tr(A\mathcal{F})\,.
\end{align}
The term $\tr(\mathcal{A}d \mathfrak{A})$ is used in the redefinition of $d B_2$. This procedure serves two purposes: it ensures $\delta_{\chi_{0}} dB_2=0$ and it fits with the dimensional reduction of $B_2$ which otherwise, due to the absence of mixed indices (between internal and 4d coordinates), does not cancel the $\tr(\mathcal{A}d \mathfrak{A})$ variation of $H_3$. The gauge anomalous variations of the \textit{universal} axion $a^\text{uni}$ cancels the one of $\Omega_3^{\text{YM},\text{4d}}$ and the gauge anomalous variations of the $\beta_r$ cancels the one of $\tr(A\mathcal{F})$. A similar analysis can be done for the Lorentz part, but as we consider a space with vanishing Ricci--tensor in the internal dimensions, those variations are not present.\\
\indent Finally, the field redefinition which ensures $dB_2$ invariance under bundle gauge transformation $\chi_{0}$, is equivalent to the analysis where the decomposition of the 10d field $B_2$ in terms of $b_2$ and $B_{1,1}$ cancels the anomaly in 4d with a variation inherited from $\delta_{\chi} B_2$. This can be seen by noting that anomalous variations of the 4d axions which cancel the 4d anomaly make $\delta H_3=0$ only if the form $\tr(\mathcal{A}d \mathfrak{A})$ as well as the analog Lorentz form are absorbed in $d B_2$. By decomposing the 10d exterior derivative $d$ as $d=d_4+d_6$, the three--form field strength variation can be written as
\begin{align}
\delta H_3=&\; \delta d_{4} b_2+[d_{4}(\tr \Theta d\omega)-d_{4}\tr(\chi dA)] +[d_{4} \delta \alpha_iR_i+d_{4}\delta \beta_r E_r]+[d_{4}(\tr \Theta d\mathcal{W})-d_{4}\tr(\chi d\mathcal{A})] \nonumber\\
&+d_{6} [\tr(\Theta d_{4} \omega)-\tr(\chi d_{4} A)]+d_{6} [\tr(\Theta d_{6} \mathcal{W})-\tr(\chi d_{6} \mathcal{A})]\,.
\end{align}
It is apparent that the second row, which can be written as $\delta \tr \mathcal{R}d\mathfrak{R}-\delta \tr \mathcal{A}d\mathfrak{A}$ has to be absorbed in the whole $d B_2$ because the index structure of its decomposition cannot cancel this variation. This is how we implement the Green--Schwarz mechanism in blow--up.

\subsection{Anomalies in the resolved space}

\begin{figure}
\centering
\includegraphics[width=0.7\textwidth]{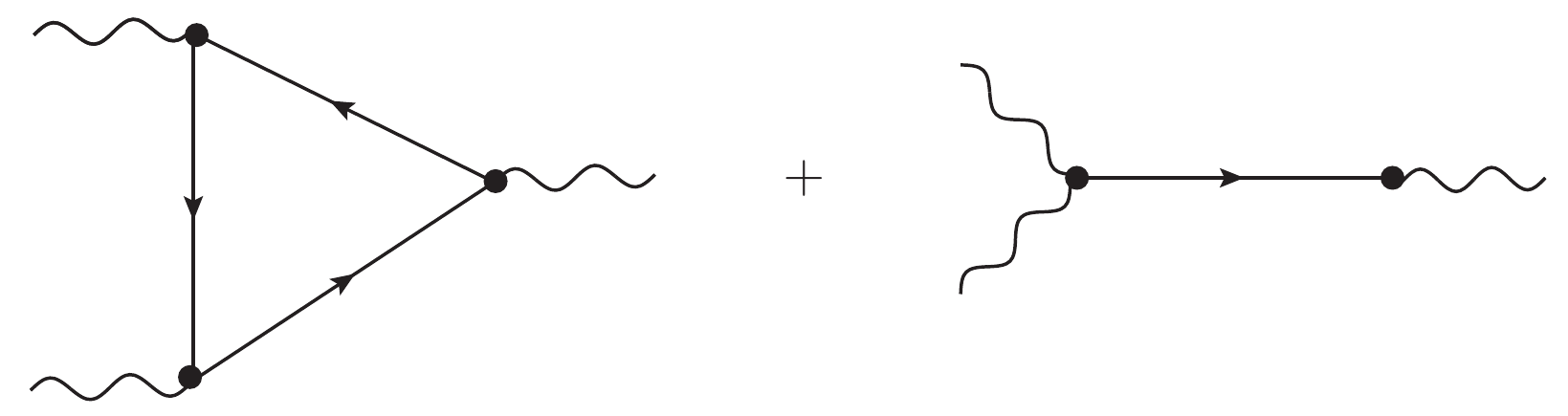}
\caption{Triangle graph inducing the gauge 4d anomalies and the axionic Green--Schwarz counter--term.}
\label{fig:anomalyGraph}
\end{figure}

Now let us proceed to the calculation of the dimensional reduction of the 10d anomaly for our explicit blow--up model. First we give a general description of every term in the 4d anomaly polynomial. Then we investigate the pure $U(1)$, $U(1)\times\text{grav}^2$ and $U(1)\times G^2$ polynomials. As the pure gravitational anomalies are canceled by the presence of $496$ gauginos in ten dimensions we do not include them in further discussions. After this we calculate the anomalies in blow--up in two different ways: from the coefficients appearing in the anomaly polynomial \eqref{eq:4d_Anomaly_Polynomial_Raw} and field--theoretically from the triangle anomaly graph given in figure \ref{fig:anomalyGraph}. The fact that both results coincide provides a non--trivial cross--check for the spectrum computation and the field redefinitions explained in section \ref{sec:SpectrumMatching}. Expanding \eqref{I12} in 6d and 4d fields, one obtains \cite{Blumenhagen:2005ga,Nibbelink:2009sp}
\begin{align}
\label{eq:4d_Anomaly_Polynomial_Raw}
I_{6} = \mathlarger{\int}\limits_{\mathcal{M}} \Big\{ \frac{1}{6}\left( \tr[\mathcal{F}^{\prime} F^{\prime}] \right)^{2} \!+\! \frac{1}{4}\left(\!\tr\mathcal{F}^{\prime 2}\!-\!\frac{1}{2}\tr\mathcal{R}^{2}\!\right)\tr F^{\prime 2} \!-\!\frac{1}{8}\left(\!\tr\mathcal{F}^{\prime 2}\!-\!\frac{5}{12}\tr\mathcal{R}^{2}\!\right)\tr R^{2} \Big\} \tr [\mathcal{F}^{\prime}F^{\prime}] \!+\! (^{\prime}\rightarrow ^{\prime\prime})\,.
\end{align}
In both $E_8$s, the whole anomaly is multiplied by a factor $\tr(\mathcal{F}F)$. This factor projects onto the $U(1)$ part of $F$, as our gauge background is by construction Abelian. In addition, it is generically only different from zero for anomalous $U(1)$s, as $T_{U(1)} \perp V_r$ for non--anomalous $U(1)$s. This means, that unless a miraculous cancellation occurs, the number of anomalous $U(1)$ is given by the rank of the $16\times 21$ matrix $V^I_r$. In our example all $U(1)$s are anomalous in blow--up, so we get contributions for all Abelian gauge group factors.

So let us discuss how the different $U(1)$ anomalies are encoded in \eqref{eq:4d_Anomaly_Polynomial_Raw} in detail:
\begin{itemize}
 \item Term 1: As $\tr(\mathcal{F}F)$ projects onto the $U(1)$-part, only pure $U(1)$ anomalies can arise from this term. The whole term contains $[\tr(\mathcal{F}F)]^3=E_r E_{r'} E_{r''} V_r^I V_{r'}^J V_{r''}^K F_I F_J F_K$. Depending on the values of $I$, $J$, and $K$, we get $U(1)^3$ anomalies if $I=J=K$, $U(1)^2 U(1)'$ anomalies if $I=J\neq K$, or $U(1) U(1)' U(1)''$ anomalies if $I\neq J\neq K\neq I$.
 \item Term 2: Here, we have a term $\tr(F)^2 \tr(\mathcal{F}F)$. The term $\tr(F)^2$ contains an inner product of the 4d field strength with itself, so from here we can get both Abelian and non--Abelian factors depending on the choice of the group element.
 \item Term 3: This term couples the 4d field strength to the 4d curvature. Hence, this term gives rise to the $U(1) \times \text{grav}^2$ anomalies.
\end{itemize}

\indent As mentioned above, the anomalies can also be evaluated in the 4d effective field--theory through the triangle Feynman graphs and counter--terms arising from couplings between axions and fermions (cf.\ figure \ref{fig:anomalyGraph}). The different anomalous contribution are given schematically by
\begin{align}
\label{SumAnomalies}
\renewcommand{\arraystretch}{1.9}
\begin{array}{l@{\,}l}
U(1)\times U(1)'\times U(1)''&:\quad \textbf{sym} \sum\limits_{\lambda} N(\lambda) (T_{U(1)}\cdot \lambda)(T_{U(1)'}\cdot \lambda)(T_{U(1)''}\cdot \lambda),\\
U(1)\times G^2&:\quad k(\mathbf{r}(G))\sum N(\mathbf{r}(G)) (T_{U(1)}\cdot (\mathbf{r}(G)))\,,\\
U(1)\times \text{grav}^2&:\quad \sum\limits_{\lambda} N(\lambda) (T_{U(1)}\cdot \lambda)\,.
\end{array}
\renewcommand{\arraystretch}{1.0}
\end{align}
Here $N(\cdot)$ denotes the multiplicity of the state in brackets and negative values indicate the conjugate representation as given by \eqref{eq:multiplicityOperator}. 
$T_{U(1) }\cdot \lambda$ represents the charge of a given \eexee\ lattice vector $\lambda$, $k(\mathbf{r})$ is the Dynkin index of the irrep and \textbf{sym} accounts for the symmetry factor corresponding to the various $U(1)$ anomalies. For the first and last terms, the sum runs over all roots, whereas for the mixed $U(1)\times G^2$ anomalies, the sum runs over the roots transforming in the respective representation only. Taking into account the numerical factors, the values of these quantities should match the coefficients of the corresponding term in the anomaly polynomial. As discussed below, we have computed both the dimensional reduction of the anomaly \eqref{I12} and the triangle anomalous graphs in the effective field theory, finding that the coefficients in (\ref{eq:4d_Anomaly_Polynomial_Raw}) coincide with the result of (\ref{SumAnomalies}). This agreement provides an important cross--check. 

In order to obtain the three different kinds of anomalies explicitly, we choose an \eexee\ Cartan basis given in \eqref{eq:CartanBasis1} in appendix \ref{app:U1Basis}, in which the eight elements $T_j,\ j=1...8$ are the $U (1)^8$ generators and the rest spans the Cartan subalgebra of the non--Abelian part of the gauge group. The $U(1)$ generators have components in both \ee's.

\subsubsection*{$\boldsymbol{U(1)\times G^2}$ anomalies}

Let us start the calculation of the anomalies with the explicit calculation of the $U(1)\times G^2$ contribution of \eqref{eq:4d_Anomaly_Polynomial_Raw} in the above basis. They are given by
\begin{align}
\label{non-Ab}
I_G=\left(25 F_1-20 F_2 -25 F_3 -4 F_4 - 66 F_5 +18 F_6+25F_7\right)\tr F'^2-F_8\tr F''^2\,.
\end{align}%
This is now compared with the anomalies $U(1)\times SU(2)^2$, $U(1)\times SU(3)^2$ and $U(1)\times SO(10)^2$ calculated from the triangle graph using the spectrum given in appendix \ref{sec:AppendixSpectrumComparison}. The field strengths for $SU(2)$ and $SU(3)$ in the visible sector are in $\tr F'^2$ and the field strength of the hidden sector $SO(10)$ is in $\tr F''^2$. The dimensional reduced anomaly polynomial coefficients and the ones computed via the traces from the anomalous triangle diagram match exactly.

\subsubsection*{$\boldsymbol{U(1)\times \text{grav}^2}$ anomalies}

When comparing the coefficients in $I_\text{grav}$ and the values of $\tr\,Q_i$ from the 4d effective spectrum for the $U(1)\times\text{grav}^2$ anomalies, we obtain again exact agreement, after the normalization factor of $-1/24$ has been taken into account in the effective field theory computation. The polynomial reads
\begin{align}
\label{grav}
 I_\text{grav}=\frac{1}{12}\left(-166 F_1 - 136 F_2 +292 F_3 + 40 F_4 + 464 F_5 -152 F_6-187 F_7 + 8 F_8\right)\tr R^2\,.
\end{align}

\subsubsection*{Pure $\boldsymbol{U(1)}$ anomalies}

Comparing the coefficients in $I_{\text{pure}}$ with the values obtained from the 4d effective spectrum we find again a perfect agreement. Note that the symmetry factors \textbf{sym} of $1/1!$ for $\tr\,Q_I Q_J Q_K$ with $I\neq J\neq K \neq I$, $1/2!$ for $\tr\,Q_I^2 Q_J$ with $I\neq J$, and $1/3!$ for $\tr\,Q_I^3$ have to be used in the 4d anomaly graph computation. The expression for the polynomial is more involved than the one of $U(1)\times G^2$ and $U(1)\times \text{grav}^2$. It is of the schematic form
\begin{align}
\label{pure}
I_{\text{pure}} = \sum a_I F_I^3 + k_{IJ} F_I^2F_J+c_{IJK}F_IF_JF_K\,.
\end{align}

\subsection*{Anomaly universality in blow--up}
As explained above, on the orbifold we have only one axion to cancel the anomalies. Anomaly freedom then requires in particular that all three kinds of anomalies are proportional such that they can all be canceled with the same axion. In blow--up, this is generically not true. However, from \eqref{eq:4d_Anomaly_Polynomial_Raw} and the discussion thereafter, it is apparent that there are still partial anomaly universalities: one can find a $U(1)$ basis where one $U(1)$ captures all gravitational anomalies, and two further $U(1)$s capture all non--Abelian anomalies of the visible and hidden sector, respectively. The rest of the $N-3$ $U(1)$s have only pure $U(1)$ anomalies.\\
\indent In order to construct such a basis, the original basis is changed to $\{\bar{F}_J\}$ as given in \eqref{eq:CartanBasis2} in appendix \ref{app:U1Basis}. After performing the base change $F_I=K_I^J \bar{F}_J$, the relevant polynomials are given by
\begin{align}
I_G&=\bar{F}_1 \left(\tr F^2_{SU(2)}+\tr F^2_{SU(3)}\right) +\bar{F}_2\tr F^2_{SO(10)}\,,\\
I_\text{grav}&=\bar{F}_3 \tr R^2\,.
\end{align}
The expression for $I_{\text{pure}}$ in terms of the new eight $U(1)$ directions is rather involved so we refrain from giving it explicitly here. While the non--Abelian $U(1)\times SU(N)^2\,,~N=2,3$ and $U(1)\times SO(10)^2$ directions are orthogonal, the $U(1)\times \text{grav}^2$ is not orthogonal to any of them. 

\subsection{Relating the anomalies on the orbifold and in blow--up}
\label{subsec:MatchingModel1}
On the orbifold there is a single anomalous Abelian gauge symmetry $U(1)_A$. This anomalous $U(1)_A$ induces an FI term which has to be canceled in a supersymmetric vacuum solution. This is done by assigning vevs to certain charged fields which in general are also charged under other $U(1)$s. Thus, once the vevs are given, we expect the breakdown of further $U(1)$s. This breakdown manifests itself from the blow--up perspective in $U(1)$s which become anomalous. The anomaly is canceled with the Green--Schwarz mechanism, which also gives a mass to the $U(1)$s. Thus we aim at investigating the 4d anomaly from the point of view of the orbifold and the blow--up. Via the descent equations, we get relations between the universal axion on the orbifold canceling the unique $U(1)_A$ anomaly and the axions in blow--up (universal and non--universal) canceling the multiple $U(1)$ anomalies.

\subsubsection*{4d anomaly from the orbifold point of view}
On the orbifold, our starting point is the anomaly polynomial $I^\text{orb}$ which describes the single unique anomalous $U(1)$ on the orbifold. To this anomaly, we add the anomaly change which is due to the departure from the orbifold point when blowing up. These changes are induced by blow--up modes that acquire vevs and thus provide mass terms via Yukawa couplings, and by the field redefinitions. We call this contribution $I^{\text{red}}$. Thus, form the orbifold perspective, the 4d anomaly polynomial $I_6$, after assigning vevs to twisted fields, decomposes as
\begin{align}
I_6=I^\text{orb}+I^\text{red}\,.
\end{align}

\subsubsection*{4d anomaly from the blow--up point of view}
In blow--up, we start from the factorized anomaly polynomial in 10 dimensions \eqref{I12}, integrate out the internal space $\mathcal{M}$, and decompose the polynomial into a universal term $I^{\text{uni}}$ plus a non--universal term $I^{\text{non}}$:
\begin{align}
I_6=I^\text{uni}+I^\text{non}=\int_\mathcal{M} X_{6,2}X_{0,4}+\int_\mathcal{M} X_{2,2} X_{4,4}\,.
\label{fact}
\end{align}
The forms $X_{2k,2l}$ were defined in section \ref{subsec:nonUniversalAxions}. The explicit decomposition of $X_4$ and $X_8$ in terms of internal and four dimensional indices is given in appendix \ref{Polynomials}. Note that the term $\int X_{2,6}X_{4,0}$ vanishes due to the Bianchi identities, and is thus not present. For later convenience, we introduce the short--hand expressions
\begin{align}
X_2^\text{uni}:=\int_{\mathcal{M}}X_{6,2}\,,\qquad X_4^\text{uni}:=X_{0,4}\,,\qquad E_r X_2^r:=\frac{1}{12}\cdot 2\,\tr(\mathcal{F}F)\,,\qquad  X_4^r:=\int_{\mathcal{M}}X_{4,4}E_r\,.
\end{align}
A factor $-1/12$ coming from the dimensional reduction is absorbed in the forms $X_2^\text{uni}$ and $X_2^r$. The expression $\int X_{6,2}X_{0,4}$ has terms mixing both $E_8$s ($'$ and $''$). This could also happen in $\int X_{4,4}X_{2,2}$. However, it turns out that these mixed terms are absent in the whole $I_6$ in \eqref{eq:4d_Anomaly_Polynomial_Raw}, which has the first and the second $E_8$ anomalies fully separated \cite{Blumenhagen:2005ga}.  

\subsubsection*{Descent equations}
Putting together the pieces described above, we obtain a relation between the anomaly polynomials on the orbifold and in blow--up:
\begin{align}
I^{\text{orb}}+I^{\text{red}}&=I^{\text{uni}}+I^{\text{non}}\,,\nonumber\\
F^{\text{orb}}X^{\text{orb}}_4+\sum\limits_a q^a_I F^I X^{\text{red}}_{4,a}&=X^{\text{uni}}_2 X^{\text{uni}}_4+\sum\limits_r X^r_2 X^r_4\,.
\label{wholepol} 
\end{align}
All the different factors in the polynomials $X^r_2,\ X^r_4,\ X^{\text{uni}}_2,\ X^{\text{uni}}_4$ are given in appendix \ref{Polynomials}. 
The counter--terms of the axions involved in the cancellation of the anomalies described above are related via the descent equation as
\begin{align}
a^{\text{orb}}X^{\text{orb}}_4+\sum_a \tau_a X^{\text{red}}_{4,a}=a^{\text{uni}}X^{\text{uni}}_4+\sum_r \beta_r X^r_4\,.
\label{descent}
\end{align}
The left hand side contains the unique orbifold axion $a^{\text{orb}}$ together with the blow--up modes $\tau_a$, and the right hand side contains the universal axion $a^{\text{uni}}$ in blow--up as well as the non--universal axions $\beta_r$. This last equation helps us to express the axions in terms of the blow--up modes. In \eqref{descent} we have added a counter--term $\sum_a \tau_a X^{\text{red}}_{4,a}$ of blow--up modes whose variation accounts for the change of the orbifold anomaly. Our aim is to express $\beta_r$ and $a^{\text{uni}}$ in terms of $a^{\text{orb}}$ and $\tau_a$, in order to confirm the interpretation of the non--universal axions as phases of the blow--up modes \cite{Nibbelink:2007ew}. We do so by calculating the four different anomalies $I^\text{orb}$, $I^\text{red}$, $I^\text{uni}$, and $I^\text{non}$ of \eqref{wholepol} separately. Then, we infer the relationship among the axions via the descent equations \eqref{descent}.

\subsubsection[Universal orbifold anomaly $I^\text{orb}$]{Universal orbifold anomaly $\boldsymbol{I^\text{orb}}$}
On the orbifold, we can choose a basis of $U(1)$ charges such that the single anomalous $U(1)_A$ is generated by

\begin{align}
T_\text{A}=\left(3,3,1,1,1,5,-3,-3,0,-4,2,0,0,0,0,0\right) 
\label{eq:U1AOrbifoldGenerator}
\end{align}
in terms of an orthogonal standard base for the Cartan elements of \eexee. With this anomalous $U(1)$ generator, the anomaly polynomial on the orbifold is
\begin{align}
\label{I6het}
I^\text{orb}=6 F_1\left(\tr F^2_{SU(2)}+\tr F^2_{SU(3)}+\tr F^2_{SO(10)}-\tr R^2 +\kappa^{IJ} \sum_{I,J}F_IF_J\right)\,.
\end{align}
The numerical factors $\kappa^{IJ}$ are not given explicitly because they are not relevant in further discussions. The factor of 6 could be absorbed by changing the normalization of $T_\text{A}$. However, we prefer not to do so, as otherwise we find this factor of 6 in all field redefinitions in the next section.

\subsubsection[Anomaly from field redefinition $I^\text{red}$]{Anomaly from field redefinition $\boldsymbol{I^\text{red}}$}
As explained in section \ref{sec:SpectrumMatching}, there is a field redefinition between the states on the orbifold and in blow--up. This field redefinition also induces a change of the anomaly polynomial described by $I^{\text{red}}$. We calculate this change by splitting up $I^{\text{red}}$ into contributions from the three types of anomalies, $I^{\text{red}}=I^{\text{red}}_{\text{G}}+I^{\text{red}}_{\text{grav}}+I^{\text{red}}_{\text{pure}}$, which we will now compute.

\subsubsection*{$\boldsymbol{U(1)\times G^2}$ anomaly redefinition}

In order to compute the redefinition of the $U(1)\times G^2$ anomaly polynomial we need to consider the change of $\tr\,Q_I$ when going from the orbifold to blow--up, where the trace is taken over the fields charged under the non--Abelian group. The change is due to the field redefinitions and to the fact that some fields become massive in blow--up and hence are not present in the massless spectrum anymore. Recall that $Q^{\gamma}_I,\ Q'^{\gamma}_I,\ \Delta^{\gamma}_I$ denote the charges of a state $\gamma$ on the orbifold, the charges in blow--up, and the shift in the charge caused by field redefinitions, see \eqref{redef}. 

The sum of the charges in blow--up $\tr\,(Q_I)_\text{BU}=\sum_{\alpha}Q^{'\alpha}_I$ runs over the states $\alpha$ that remain massless after giving vevs to the blow--up modes. Hence, in order to recover the trace on the orbifold prior to having assigned vevs, we also have to include a sum over the states that gain a mass in blow--up, which we label by $\beta$. We thus obtain
\begin{align}
\tr(Q_I)_\text{BU}&=\sum\limits_{\alpha}Q^{\alpha}_I-\sum\limits_{\alpha}\Delta^{\alpha}_I
=\sum\limits_{\alpha}Q^{\alpha}_I-\sum\limits_{\alpha}\Delta^{\alpha}_I+\sum\limits_{\beta}Q^{\beta}_I-\sum\limits_{\beta}\Delta^{\beta}_I-\sum\limits_{\beta}Q^{'\beta}_I\nonumber\\
&=\tr(Q_I)_\text{orb}-\sum\limits_{\gamma=\alpha,\beta}\Delta^{\gamma}_I-\sum\limits_{\beta}Q^{'\beta}_I\,,
\end{align}
where we added a $0$ in the first step and rearranged the terms in the second step. Note that the last sum $\sum_\beta Q'^\beta_I$ which sums over all fields that became massive in blow--up, vanishes identically: all massive states are vector--like with respect to their charges, so the sum always contains pairs of opposite charges. Leaving out this last term, the contribution to the 4d anomaly polynomial and the redefinition part read
\begin{align}
I_{G}&= F^I \tr F^2_G \sum_{\alpha} Q'^{\alpha}_I\,,\nonumber\\
I^\text{red}_{G}&\sim \sum_{G,I}\left(-\sum_{\gamma}\Delta^{\gamma}_I \right)F^I \tr F_G^2 \sim \sum_{G,I} c^{G}_IF^I \tr F_G^2\,.
\label{red1}
\end{align}
In the sums $G$ runs over \SU{2}, \SU{3} and \SO{10}. When evaluating the sum and comparing with the orbifold result, we obtain a perfect match of all $U(1)\times G^2$ anomalies of both theories. The anomaly coefficients $c^{G}_I$ of \eqref{red1} are given by
\begin{align}
c^{SU(2),SU(3)}_I&=\left(19, -20,-25,-4,-66,18,25,0\right)\,,\nonumber\\
c^{SO(10)}_I&=\left(-6,0, 0, 0, 0, 0, 0, -1\right)\,.
\end{align}

\subsubsection*{$\boldsymbol{U(1)\times\text{grav}^2}$ anomaly redefinition}

For the $U(1)\times\text{grav}^2$ anomaly one has to include all the massless fields in the trace. This means that, in contrast to the $U(1)\times G^2$ anomalies, one also has to add the contribution coming from the Abelian blow--up mode charges $q^a_I$. The the contribution to the 4d anomaly polynomial and the redefinition part is then given by 
\begin{align}
I_{\rm grav}&\sim F^I \tr R^2 \tr (Q'_I)_{\rm BU}= F^I\tr R^2 \sum_{\alpha} Q'^{\alpha}_I\nonumber\\
&= F^I\tr R^2 \left(\sum_{\alpha} Q^{\alpha}_I-\sum_{\alpha}\Delta^{\alpha}_I+\sum_{\beta}Q^{\beta}_I-\sum_{\beta}\Delta^{\beta}_I-\sum_{\beta}Q^{'\beta}_I+\sum_a q^a_I-\sum_a q^a_I\right)\,,\nonumber\\
I^{\rm red}_{\rm grav}&\sim \left(-\sum_{\gamma=\alpha,\beta}\Delta^{\gamma}_I-\sum_a q^a_I\right)F^I\tr R^2= c_I^\text{grav} F^I\tr R^2\,,\label{red2}
\end{align}
where we again added the contributions from the massive fields and used that $\sum_\beta Q'^\beta_I=0$. The index $\gamma$ contains both $\alpha$ for massless and $\beta$ for massive fields. The anomaly coefficients in \eqref{red2} are 
\begin{align}
c_I^\text{grav}=\left(-\frac{47}{6}, -\frac{34}{3}, \frac{73}{3}, \frac{10}{3}, \frac{116}{3}, -\frac{38}{3}, -\frac{187}{12}, \frac{2}{3}\right)\,.
\end{align}
We find again a perfect match between the blow--up polynomial and the redefined one, supporting the obtained field redefinitions \eqref{eq:FieldRedef}.

\subsubsection*{Pure $\boldsymbol{U(1)}$ anomaly redefinition}
A similar procedure can be applied to the pure $U(1)$ anomalies and in this case the field redefinitions change the polynomial via
\begin{align}
I_{\rm pure}&\sim \frac{1}{3!}\sum_{I,J,K} F^I F^J F^K\sum_{\alpha}Q'^{\alpha}_IQ'^{\alpha}_JQ'^{\alpha}_K \nonumber\\
&= \frac{1}{3!}\sum_{I,J,K} F^I F^J F^K\left(\sum_{\alpha}Q^{\alpha}_IQ^{\alpha}_JQ^{\alpha}_K+ \sum_{a}q^{a}_Iq^{a}_Jq^{a}_K+\sum_{\beta}Q^{\beta}_IQ^{\beta}_JQ^{\beta}_K\right)+I^\text{red}_\text{pure}\nonumber\\
&= \frac{1}{3!}\sum_{I,J,K} F^I F^J F^K\tr(Q_IQ_JQ_K)_\text{orb}+I^\text{red}_\text{pure}\,,\nonumber\\
I^{\rm red}_{\rm pure}&\sim \frac{1}{3!}\sum_{I,J,K} F^I F^J F^K \left( \phantom{-}\sum_{\gamma=\alpha,\beta}(-3\Delta^{\gamma}_I Q^{\gamma}_JQ^{\gamma}_K +3 \Delta^{\gamma}_I \Delta^{\gamma}_JQ^{\gamma}_K - \sum_{\gamma=\alpha,\beta}\Delta^{\gamma}_I\Delta^{\gamma}_J \Delta^{\gamma}_K\right.\nonumber\\ &\qquad\qquad\qquad\qquad\qquad\!\!\!\left.-\sum_{a}q^{a}_Iq^{a}_J q^{a}_K-\sum_{\beta}Q'^{\beta}_IQ'^{\beta}_JQ'^{\beta}_K \right)\,.\label{redef4}
\end{align}
We have made explicit a factor of $1/3!$ coming from the symmetry factor \textbf{sym} and from permutation symmetries of the sum.
The anomalies match perfectly when assuming the mass terms to have the structure explained in section \ref{sec:SpectrumMatching}. The coefficients of the anomaly terms turn out to be rather big. For example, the coefficients of the cubic anomaly term $\sum_I c_I^\text{pure} F_I^3$ are given by 
\begin{align}
 c_I^\text{pure}=\frac{1}{3!}\left(14576, 91184, -436928, -202064, -384592, 270832, 24026, -16\right)\,.
\end{align}
The expression for $I^{\text{red}}$ simplifies due to the fact that $\sum_{\beta}Q'^{\beta}_IQ'^{\beta}_JQ'^{\beta}_K=0$ and $\sum_{\beta} Q'^{\beta}_I=0$, thus we obtain 
\begin{align}
I^{\text{red}}\;=\;&-\sum_a q^a_I F^I\left( \sum_{\gamma=\alpha,\beta}r^{\gamma}_a\tr F^2_{\text{G}}
+(1+\sum_{\gamma=\alpha,\beta}r^{\gamma}_a)\tr R^2\right.\nonumber\\
&\;\left.+\frac{1}{3!}F^J F^K \left[
 3 \sum_{\gamma}r^{\gamma}_a Q^{\gamma}_J Q^{\gamma}_K-3 \sum_{\gamma}r^{\gamma}_a \Delta^{\gamma}_J Q^{\gamma}_K + \sum_{\gamma}r^{\gamma}_a \Delta^{\gamma}_J\Delta^{\gamma}_K +q^a_Jq^a_K \right]\right)\,.
\label{notrue}
\end{align}
In the sum running over $a=(k,\sigma)$, the factors $r^{\gamma}_a$ not appearing in \eqref{redef} are zero.

\subsubsection[Universal blow--up anomaly $I^\text{uni}$]{Universal blow--up anomaly $\boldsymbol{I^\text{uni}}$}
The universal anomaly in blow--up is given by 
\begin{align}
I^\text{uni} &= \int\limits_\mathcal{M} X_{2}^\text{uni} X_4^\text{uni}\nonumber\\
&= -\frac{1}{12} \int\limits_\mathcal{M} (\tr R^2-\tr F^2)\left(\text{tr}(\mathcal{F}'F')\text{tr}\mathcal{F}'^2-\frac{1}{2}\text{tr}\mathcal{F}'^2\text{tr}(\mathcal{F}''F'') -\frac{1}{4}\text{tr}(\mathcal{F}'F')\text{tr}\mathcal{R}^2 +'\leftrightarrow''\right)\,.
\end{align}
Using the intersection numbers and the expansion of the internal flux $\mathcal{F}$, we obtain for the universal anomaly in blow--up
\begin{align}
I^\text{uni} & = \frac{1}{2}(\tr R^2-\tr F^2)\cdot\left(-25 F_1 +20 F_2 + 25 F_3 + 4 F_4 + 66 F_5- 18 F_6 - 25 F_7 - F_8)\right)\,.
\end{align}

\subsubsection[Non--universal local anomalies $I^\text{non}$]{Non--universal local anomalies $\boldsymbol{I^\text{non}}$}
Lastly, we have the non--universal axions $\beta_r$ to cancel the other $U(1)$ anomalies. Their contributions are given by
\begin{align}
\label{Inon}
I^{\text{non}} &=\int\limits_\mathcal{M}X^r_2 X^r_4\,.
\end{align}
This expression is evaluated by using the Bianchi identities to express $\tr\mathcal{R}^2$ in terms of $\tr\mathcal{F}^2$ as
\begin{align}
\int\limits_{E_r} \tr \mathcal{R}^2=\int\limits_{E_r} \tr \mathcal{F}^2 =V^I_{r_1}V^I_{r_2}E_{r_1}E_{r_2}E_r\,.
\label{eq:BIOverE}
\end{align}
In appendix \ref{Polynomials} the expressions for  $X^r_4$ and $X^r_2$ are given. The integration in (\ref{Inon}) is performed by using the intersection numbers.  We obtain
\begin{align}
I^{\text{non}}\;=\;&\frac{1}{2}(-25 F_1+20 F_2 +25 F_3-4 F_4- 66 F_5+ 18 F_6 + 25 F_7+ F_8)\nonumber\\
 &\;\cdot \left( \tr F^2_{SO(10)} -\tr F^2_{SU(2)}- \tr F^2_{SU(3)}\right )+\sum_{IJK} h^{IJK}F_IF_JF_K\nonumber\\
 &\;+ \frac{1}{12}(-16 F_1 - 256 F_2 + 142 F_3+ 16 F_4 + 68 F_5 - 44 F_6 - 37 F_7 + 2 F_8)\tr R^2\,,
\end{align}
where we have expressed the coefficients corresponding to pure $U(1)$ anomalies schematically as $h^{IJK}$.  Now we have computed all 4 contributions to the anomalies in \eqref{wholepol}.

\subsection{Relation among the axions}

From the above results for $I^\text{orb}$, $I^\text{red}$, $I^\text{uni}$, and $I^\text{non}$, we can now establish the relation between the single orbifold axion, the axions in blow--up, and the blow--up modes using the descent equations \eqref{descent}. We need to make an ansatz to factorize $I^{\text{red}}$ which is compatible with this interpretation. A given factorization  $I^{\text{red}}=\sum_a q^a_IF_IX_{4,a}^{\text{red}}$ is canceled via the counter--term $\sum_a \tau_{a} X_{4,a}^{\text{red}}$. The indices $a$ and $r$ run over the same set, so we use only $r$. Considering $X^{\text{orb}}_4= -6X^{\rm uni}_4$ we make the following ansatz for relating the various axions
\begin{align}
\beta_r=d_r \tau_r\,, \qquad a^{\rm uni}=-6a^{\text{orb}}+\sum\limits_r c_r \tau_r\,.
\label{eq:AxionAnsatz}
\end{align}
Here, the $c_r$ and $d_r$ are coefficients in the linear combinations and the factor of $-6$ arises due to the normalization choice in \eqref{eq:U1AOrbifoldGenerator}. Substituting this ansatz into \eqref{descent}, the 4--form involved in the factorization is expressed as
\begin{align}
X^{\text{red},r}_4=c_r X^{\rm uni}_4+d_rX^r_4\,.
\end{align}
Substituting this last expression into $I^{\text{red}}$ in \eqref{wholepol} yields
\begin{align}
\label{ansatz}
I^{\rm red}&=\sum_r q^r_IF_I \left(c_r X^{\rm uni}_4 +d_r X^r_4\right)\,.
\end{align} 
Looking at the whole anomaly polynomial (\ref{wholepol}), we impose equality of each factor on the left hand side and on the right hand side. As there are 8 anomalous $U(1)$s, we obtain 152 equations in total, where 8 equations arise from the 8 $U(1)\times\text{grav}^2$ anomalies, $8\cdot 3=24$ equations arise from the mixed $U(1)\times G^2$ anomalies, and $8+8\cdot7+8\cdot7\cdot6/3!=120$ equations arise from the pure $U(1)$ anomalies. At first sight, this system is highly over--constrained, as we only have $2\cdot21=42$ coefficients $c_r,d_r$. However, as it turns out, only 29 out of the 152 equations are independent. In particular, we find that part of the solution is $d_r=-1/6$ for all $r$. The factor of 6 arises again due to our normalization convention. From \eqref{eq:AxionAnsatz} we thus see that  axions $\tau_r$ coming from field redefinitions are indeed the same as the non--universal axions $\beta_r$, which are responsible for canceling the non--universal anomalies in blow--up. This result allows us to interpret the blow--up modes as non--universal axions in a compact resolution of the \zseven\ orbifold.

However, choosing a common value for all $c_r$ or grouping them by fixed points or by sectors turns out to be impossible. This implies that the universal axion in blow--up is a mixture of the unique orbifold axion and the blow--up modes.

\section{Conclusion}
\label{sec:Conclusion}
The analysis of the paper shows that a careful inspection of the blow--up mechanism reveals detailed information about the models away from the orbifold point. With the concept of local multiplicity operators the knowledge about orbifold properties can be carried over to the blow--up model. Within the framework of our \zseven\ example we can study the match of the spectrum in detail. All relevant states can be identified on both sides. Masses can be compared and some subtleties (concerning masses in the large volume limit) can be clarified. 

We have emphasized that the study of the Green--Schwarz anomaly polynomial is a key tool to understand the resolution of the orbifold point. In contrast to the single $U(1)_A$ of the orbifold model we find many anomalous  $U(1)$s in blow--up and we identify the corresponding localized axions. Mixing of the axion in the anomaly polynominal is relevant for the interactions in the blow--up model. The match with the anomalies supports the reliability of the field--theoretical methods used in the resolution procedure.

Our analysis shows that it pays off to study the blow--up mechanism in detail. It allows us to carry over the powerful computational techniques of orbifold compactification to smooth compactifications (where otherwise only effective field theory methods in the large volume limit are available). Here we have employed an example based on the \zseven\ orbifold which shares the complexity of realistic models but avoids some of the subtleties found e.g.\ in the models of the Mini--Landscape. These subtleties are not yet completely understood, but they seem to be no obstructions in principle. We hope that with the methods developed here these problems can be overcome.

\section*{Acknowledgments}
We thank Michele Trapletti for collaboration at early stages of this project. We would also like to thank Damian Mayorga Pe\~{n}a, Stefan Groot Nibbelink and Ivonne Zavala for useful discussions. NGCB would like to thank the ``Centro de Aplicaciones Tecnol\'ogicas y Desarrollo Nuclear'' (CEADEN) and ``Proyecto Nacional de Ciencias B\'{a}sicas Part\'{i}culas y Campos''(CITMA, Cuba).  The work was partially supported by the SFB-Tansregio TR33 ``The Dark Universe'' (Deutsche Forschungsgemeinschaft) and the European Union 7th network program ``Unification in the LHC era'' (PITN-GA-2009-237920).

\clearpage
\appendix

\section{Orbifold and blow--up spectrum}
\label{sec:AppendixSpectrumComparison}
This appendix contains a detailed list of all orbifold and blow--up states. For each state the local and global multiplicity is given, as well as the characteristic data (i.e.\ the \eexee\ roots for the blow--up states and the shifted momenta for the orbifold states) together with the field redefinition between these states. The organization of the table is as follows: it is divided into blocks where each block corresponds to an \eexee\ roots in blow--up. Below this root, we list all orbifold states which are redefined to this root, where the redefinition used is indicated in the last column.\\ 
We give the representation (of the blow--up root) or an auxiliary name (for the orbifold states) in the first column. The second column contains the twisted sector where the orbifold state lives (for the blow--up states this information is not defined anymore). The entry $1$-$7$ indicates an untwisted state. The third column gives the local multiplicity, i.e.\ the multiplicity of each state at each fixed point. The ``tot'' column contains the total multiplicity, i.e.\ the sum of the local multiplicities over all fixed points. In our convention, we list only the highest states of non--Abelian irreps, where a negative multiplicity indicates that the state belongs to the complex conjugate representation. The last block of the table contain the 21 orbifold states which were chosen as blow--up modes.
 \renewcommand{\arraystretch}{1.2}
 \scriptsize
 \enlargethispage{\baselineskip}
\begin{longtable}{|c|c|c|c|c|c|c|c|c|c|l|c|}
 \hline
 \multirow{2}{*}{State} & \multirow{2}{*}{Sector} & \multicolumn{7}{c|}{Local multiplicity} & \multirow{2}{*}{tot} & \multirow{2}{*}{\eexee\ root / $P_{\text{Sh}}$} & \multirow{2}{*}{Redef}\\
 \cline{3-9}
 			&			  & 1&2&3&4&5&6&7			   &			& &      \\
 \hline
 \hline
\endhead
%
%
  (3,2,1)&--&$\frac{1}{7}$&$\frac{1}{7}$&$\frac{1}{7}$&$\frac{1}{7}$&$\frac{1}{7}$&$\frac{1}{7}$&$\frac{1}{7}$&$1$&$\big(1$,$0$,$0$,$0$,-$1$,$0$,$0$,$0\big)\big(0$,$0$,$0$,$0$,$0$,$0$,$0$,$0$\big)&--\\ 
 \hline
  $Q_1$&1-7&$\frac{1}{7}$&$\frac{1}{7}$&$\frac{1}{7}$&$\frac{1}{7}$&$\frac{1}{7}$&$\frac{1}{7}$&$\frac{1}{7}$&$1$&$\big(1$,$0$,$0$,$0$,-$1$,$0$,$0$,$0\big)\big(0$,$0$,$0$,$0$,$0$,$0$,$0$,$0$\big)&none\\
 \hline
 \hline
  (3,2,1)&--&$1$&-$\frac{1}{7}$&-$\frac{1}{7}$&$\frac{1}{7}$&-$\frac{1}{7}$&$\frac{1}{7}$&$\frac{1}{7}$&$1$&$\big(\frac{1}{2}$,-$\frac{1}{2}$,$\frac{1}{2}$,$\frac{1}{2}$,-$\frac{1}{2}$,$\frac{1}{2}$,-$\frac{1}{2}$,-$\frac{1}{2}\big)\big(0$,$0$,$0$,$0$,$0$,$0$,$0$,$0$\big)&--\\ 
 \hline
  $Q_2$&2&$1$&0&0&0&0&0&0&$1$&$\big(\frac{1}{2}$,-$\frac{1}{2}$,$\frac{3}{14}$,$\frac{3}{14}$,-$\frac{11}{14}$,-$\frac{1}{14}$,-$\frac{1}{14}$,$\frac{3}{14}\big)\big(\text{-}\frac{2}{7}$,-$\frac{2}{7}$,$0$,$0$,$0$,$0$,$0$,$0$\big)&\eqref{eq:redef1}\\
 \hline
 \hline
  (3,2,1)&--&$1$&-$\frac{1}{7}$&-$\frac{1}{7}$&$\frac{1}{7}$&-$\frac{1}{7}$&$\frac{1}{7}$&$\frac{1}{7}$&$1$&$\big(\frac{1}{2}$,-$\frac{1}{2}$,$\frac{1}{2}$,$\frac{1}{2}$,-$\frac{1}{2}$,-$\frac{1}{2}$,$\frac{1}{2}$,-$\frac{1}{2}\big)\big(0$,$0$,$0$,$0$,$0$,$0$,$0$,$0$\big)&--\\ 
 \hline
  $Q_3$&1&$1$&0&0&0&0&0&0&$1$&$\big(\frac{1}{2}$,-$\frac{1}{2}$,$\frac{5}{14}$,$\frac{5}{14}$,-$\frac{9}{14}$,$\frac{3}{14}$,$\frac{3}{14}$,$\frac{5}{14}\big)\big(\text{-}\frac{1}{7}$,-$\frac{1}{7}$,$0$,$0$,$0$,$0$,$0$,$0$\big)&\eqref{eq:redef1}\\
 \hline
 \hline
%
%
%
  ($\overline{3}$,1,1)&--&$\frac{1}{7}$&$\frac{1}{7}$&-$\frac{1}{7}$&$\frac{1}{7}$&-$\frac{1}{7}$&-$\frac{1}{7}$&-$1$&-$1$&$\big(\frac{1}{2}$,$\frac{1}{2}$,-$\frac{1}{2}$,-$\frac{1}{2}$,$\frac{1}{2}$,$\frac{1}{2}$,$\frac{1}{2}$,$\frac{1}{2}\big)\big(0$,$0$,$0$,$0$,$0$,$0$,$0$,$0$\big)&--\\
\hline
\end{longtable}

\section{$\boldsymbol{U(1)}$ bases}
\label{app:U1Basis}
\normalsize
We use two different Cartan bases in the paper. In the first basis, the anomalous direction on the orbifold is singled out. In the second basis, the gravity and non--Abelian anomalies in blow--up are singled out.\\
\indent In the first basis, the components of the Cartan subalgebra are chosen such that the first row corresponds to the anomalous $U(1)$ generator \eqref{eq:U1AOrbifoldGenerator}. The next 7 rows correspond to other $U(1)$ generators perpendicular to $U(1)_A$. The last 8 rows are the Cartan basis of the non--Abelian group factors $\SU{3}\times\SU{2}\times\SO{10}$. The basis is given as $T_K=Q_K^I H_I$, where the $H_I$ form an orthogonal basis for \eexee\ fulfilling $\tr(H_IH_J)=\delta_{IJ}$. The matrix $Q$ reads
\begin{align}
Q_K^I=\left(
\begin{array}{cccccccccccccccc}
 3 & 3 & 1 & 1 & 1 & 5 & -3 & -3 & 0 & -4 & 2 & 0 & 0 & 0 & 0 & 0 \\
 -15 & -15 & -5 & -5 & -5 & 59 & 15 & 15 & 0 & 20 & -10 & 0 & 0 & 0 & 0 & 0 \\
 -3 & -3 & -1 & -1 & -1 & -5 & 3 & 3 & 0 & 4 & 40 & 0 & 0 & 0 & 0 & 0 \\
 -3 & -3 & 27 & 27 & 27 & -5 & 3 & 3 & 0 & 4 & -2 & 0 & 0 & 0 & 0 & 0 \\
 3 & 3 & 1 & 1 & 1 & 5 & -3 & 25 & 0 & -4 & 2 & 0 & 0 & 0 & 0 & 0 \\
 3 & 3 & 1 & 1 & 1 & 5 & 25 & -3 & 0 & -4 & 2 & 0 & 0 & 0 & 0 & 0 \\
 3 & 3 & 1 & 1 & 1 & 5 & -3 & -3 & 0 & 17 & 2 & 0 & 0 & 0 & 0 & 0 \\
 0 & 0 & 0 & 0 & 0 & 0 & 0 & 0 & 1 & 0 & 0 & 0 & 0 & 0 & 0 & 0 \\
 0 & 0 & 0 & 1 & -1 & 0 & 0 & 0 & 0 & 0 & 0 & 0 & 0 & 0 & 0 & 0 \\
 0 & 0 & 1 & -1 & 0 & 0 & 0 & 0 & 0 & 0 & 0 & 0 & 0 & 0 & 0 & 0 \\
 1 & -1 & 0 & 0 & 0 & 0 & 0 & 0 & 0 & 0 & 0 & 0 & 0 & 0 & 0 & 0 \\
 0 & 0 & 0 & 0 & 0 & 0 & 0 & 0 & 0 & 0 & 0 & 1 & -1 & 0 & 0 & 0 \\
 0 & 0 & 0 & 0 & 0 & 0 & 0 & 0 & 0 & 0 & 0 & 0 & 1 & -1 & 0 & 0 \\
 0 & 0 & 0 & 0 & 0 & 0 & 0 & 0 & 0 & 0 & 0 & 0 & 0 & 1 & -1 & 0 \\
 0 & 0 & 0 & 0 & 0 & 0 & 0 & 0 & 0 & 0 & 0 & 0 & 0 & 0 & 1 & 1 \\
 0 & 0 & 0 & 0 & 0 & 0 & 0 & 0 & 0 & 0 & 0 & 0 & 0 & 0 & 1 & -1
\end{array}
\right)\,.
\label{eq:CartanBasis1}
\end{align}
\indent The second basis for the eight $U(1)$s in Section \ref{sec:Anomalies} is given by generators $T_I (K^{-1})^I_J$. The field strengths
are related via $\bar{F}_I=K_I^JF_J$ . The matrix $K$ is given by
\begin{align}
K_I^J=\left(
\begin{array}{cccccccccccccccc}
 25 & -20 & -25 & -4 & -66 & 18 & 25 & 0 & 0 & 0 & 0 & 0 & 0 & 0 & 0 & 0 \\
 0 & 0 & 0 & 0 & 0 & 0 & 0 & -1 & 0 & 0 & 0 & 0 & 0 & 0 & 0 & 0 \\
 -\frac{83}{6} & -\frac{34}{3} & \frac{73}{3} & \frac{10}{3} & \frac{116}{3} & -\frac{38}{3} & -\frac{187}{12} & \frac{2}{3} & 0 & 0 & 0 & 0 & 0
& 0 & 0 & 0 \\
 1 & 0 & -2 & \frac{75}{4} & 0 & 0 & 0 & 0 & 0 & 0 & 0 & 0 & 0 & 0 & 0 & 0 \\
 0 & 1 & 8 & -55 & 0 & 0 & 0 & 0 & 0 & 0 & 0 & 0 & 0 & 0 & 0 & 0 \\
 0 & 0 & \frac{14}{3} & -\frac{137}{3} & 1 & 0 & 0 & 0 & 0 & 0 & 0 & 0 & 0 & 0 & 0 & 0 \\
 0 & 0 & -\frac{2}{3} & \frac{26}{3} & 0 & 1 & 0 & 0 & 0 & 0 & 0 & 0 & 0 & 0 & 0 & 0 \\
 0 & 0 & -\frac{3}{2} & \frac{125}{8} & 0 & 0 & 1 & 0 & 0 & 0 & 0 & 0 & 0 & 0 & 0 & 0 \\
 0 & 0 & 0 & 0 & 0 & 0 & 0 & 0 & 1 & 0 & 0 & 0 & 0 & 0 & 0 & 0 \\
 0 & 0 & 0 & 0 & 0 & 0 & 0 & 0 & 0 & 1 & 0 & 0 & 0 & 0 & 0 & 0 \\
 0 & 0 & 0 & 0 & 0 & 0 & 0 & 0 & 0 & 0 & 1 & 0 & 0 & 0 & 0 & 0 \\
 0 & 0 & 0 & 0 & 0 & 0 & 0 & 0 & 0 & 0 & 0 & 1 & 0 & 0 & 0 & 0 \\
 0 & 0 & 0 & 0 & 0 & 0 & 0 & 0 & 0 & 0 & 0 & 0 & 1 & 0 & 0 & 0 \\
 0 & 0 & 0 & 0 & 0 & 0 & 0 & 0 & 0 & 0 & 0 & 0 & 0 & 1 & 0 & 0 \\
 0 & 0 & 0 & 0 & 0 & 0 & 0 & 0 & 0 & 0 & 0 & 0 & 0 & 0 & 1 & 0 \\
 0 & 0 & 0 & 0 & 0 & 0 & 0 & 0 & 0 & 0 & 0 & 0 & 0 & 0 & 0 & 1
\end{array}
\right)\,.
\label{eq:CartanBasis2}
\end{align}

\section{Polynomials}
\label{Polynomials}

To obtain the factorization of the polynomial, one has to start with the expressions
\begin{align}
X_{6,2}\;=\;&-\frac{1}{12}\left[\frac{3}{2}\tr(F'\mathcal{F}')\tr \mathcal{F}'^2-\frac{3}{4}\tr(F'\mathcal{F}')\tr\mathcal{R}^2+'\leftrightarrow''\right]\\
X_{4,4}\;=\;&\tr (\mathcal{F}'F')^2 +\frac{3}{4}\tr F'^2\tr \mathcal{F}'^2-\frac{3}{8}\tr F'^2\tr \mathcal{R}^2-\frac{1}{8}\tr \mathcal{F}'^2\tr R^2+'\leftrightarrow ''\nonumber\\
&+\frac{1}{16}\tr R^2\tr \mathcal{R}^2-\tr (\mathcal{F}'F')\tr (\mathcal{F}'' F'')\,.
\end{align}
The anomaly polynomial factorization is given in terms of the following 2-- and 4--forms
\begin{align}
X^{\text{uni}}_4\;=\;&X_{0,4}=(\tr R^2-\tr F^2)\,,\\
X^r_4\;=\;&\int\limits_\mathcal{M} E_{r_1}E_{r_2}E_r \left(V^{I'}_{r_1}V^{J'}_{r_2}F'_{I'}F'_{J'}+V^{I'}_{r_1}V^{I'}_{r_2}\left(\frac{3}{4}\tr F'^2-\frac{1}{8}\tr R^2\right)+ '\leftrightarrow'' - V^{I'}_{r_1}V^{I''}_{r_2}F_{I'}F_{I''}\right) \nonumber\\
&\;+ \int\limits_\mathcal{M} \tr \mathcal{R}^2 E_r \left( \frac{1}{16}\tr R^2-\frac{3}{8}\tr F'^2-\frac{3}{8}\tr F''^2\right)\,.
\end{align}
Using the Bianchi identities \eqref{eq:BIOverE} we obtain
\begin{align}
X^r_4\;=&\;\int\limits_\mathcal{M} E_{r_1}E_{r_2}E_r \left(V^{I'}_{r_1}V^{J'}_{r_2}F'_{I'}F'_{J'}+V^{I'}_{r_1}V^{I'}_{r_2}\left(\frac{3}{4}\tr F'^2-\frac{1}{8}\tr R^2\right)+ '\leftrightarrow'' - V^{I'}_{r_1}V^{I''}_{r_2}F_{I'}F_{I''}\right)
\nonumber\\
&\;+ \int\limits_\mathcal{M} E_{r_1}E_{r_2}E_r V^I_{r_1}V^I_{r_2} \left( \frac{1}{16}\tr R^2-\frac{3}{8}\tr F'^2-\frac{3}{8}\tr F''^2\right)\,.
\end{align}
The 2--forms are given by
\begin{align}
X^{\text{uni}}_2&=\int\limits_\mathcal{M} X_{6,2}=-\frac{1}{12}\int\limits_\mathcal{M}(\text{tr}(\mathcal{F}'F')\text{tr}\mathcal{F}'^2-\frac{1}{2}\text{tr}\mathcal{F}'^2\text{tr}(\mathcal{F}''F'')
-\frac{1}{4}\text{tr}(\mathcal{F}'F')\text{tr}\mathcal{R}^2 +'\leftrightarrow'')\,,\\
X^r_2&=\frac{1}{12} V_r^IF_I\,.
\end{align}

\vskip 1cm
\bibliographystyle{paper}
\small{
\providecommand{\href}[2]{#2}\begingroup\raggedright\endgroup

}

\end{document}